\documentclass[]{rsos}

\usepackage{url}

\newcommand{\showfigures}[1]{#1} 
\newcommand{\modif}[1]{\textcolor{black}{#1}} 

\graphicspath{{Figures/}}

\begin{document}

\title{Collective cell migration without proliferation: density determines cell velocity and wave velocity}

\author{Sham Tlili,$^{1,2}$ Estelle Gauquelin,$^3$ Brigitte Li,$^1$  Olivier Cardoso,$^1$ Beno\^{\i}t Ladoux,$^{2,3}$ H\'{e}l\`{e}ne Delano\"e-Ayari$^4$ and  
Fran\c cois Graner$^1$}

\address{$^1$Laboratoire Mati\`ere et Syst\`emes Complexes, Universit\'e Denis Diderot - Paris 7, CNRS UMR 7057, Condorcet building, 10 rue Alice Domon et L\'eonie Duquet, F-75205 Paris Cedex 13, France; 
\\ 
$^2$Mechanobiology Institute Department of Biological Sciences, National University of Singapore, 5A Engineering Drive, 1 Singapore 117411; 
\\
$^3$ Institut Jacques Monod, Universit\'e Denis Diderot - Paris 7, CNRS UMR  7592,  Buffon building, 15 rue H\'el\`ene Brion, F-75205 Paris Cedex 13, France; 
\\
$^4$ Univ. Lyon, Universit\'e Claude Bernard Lyon 1, CNRS UMR 5306,  Institut Lumi\`ere Mati\`ere, Campus LyonTech - La Doua,  Kastler building, 10 rue Ada Byron,  
F-69622 Villeurbanne Cedex, France
}
\subject{biophysics, cellular biophysics, wave motion}


\corres{S. Tlili, F. Graner\\
\email{mbitlil@nus.edu.sg, francois.graner@univ-paris-diderot.fr}}

\begin{abstract} 
Collective cell migration contributes to \modif{embryogenesis}, wound healing and tumor metastasis. \modif{C}ell monolayer migration \modif{experiments help} understand\modif{ing} what determines the movement of cells far from the leading edge. 
\modif{Inhibiting cell proliferation limits cell density increase and prevents jamming; we observe long-duration migration
and  quantify} space-time characteristics of the velocity profile  over large length- and time-scales. Velocity waves propagate backwards and their frequency depends only on cell density at the \modif{moving} front. Both  cell average velocity and  wave velocity increase linearly with the cell effective radius regardless of the distance \modif{to the} front. Inhibiting lamellipodia \modif{decreases} cell velocity while waves either disappear or \modif{have a lower frequency. Our} model combines conservation laws, monolayer mechanical properties and a phenomenological coupling between strain and polarity: advancing cells pull on their followers which then become polarized. \modif{With reasona-ble values of parameters, t}his model agrees with \modif{several of} our experimental observations. Together, our experiments and model \modif{disantangle the respective contributions of active velocity and of proliferation in monolayer migration}, explain how cells maintain their polarity far from the \modif{moving} front, and highlight the importance of strain-polarity coupling and density in long-range information propagation.
\end{abstract}


\begin{fmtext}
\end{fmtext}


\maketitle

\section{Introduction}

Collective migration of cells connected by cell-cell adhesion occurs across \modif{several} time scales and length scales  in numerous biological processes like embryogenesis (notably gastrulation), wound healing, regeneration or tumor metastasis \cite{Keller2002,Friedl2004,Friedl2009,Arboleda2010}.
To study such long-range information propagation mediated by mechanical stress in tissues, \textit{in vitro} reconstructed assemblies of cohesive cells are useful experimental model systems
\cite{Trepat2009,duRoure2005} where each individual cell can grow, divide, die, migrate. 
In two-dimensional (2D) monolayers, cells interact with each other biochemically and mechanically, for instance through adhesion, and have a richer migration behaviour than single cells. 
It is possible to constrain geometrically and reproducibly control their collective migration. 
Patterned substrate of adhesive strips  enable to investigate the tissue global response to active processes \modif{such as} cell migration \cite{Trepat2009,Angelini2010}  or cell division  \cite{Puliafito2012}, and quantitatively test the  impact of drugs \modif{like} blebbistatin  \cite{Vedula2012}. 
Madin-Darby canine kidney (MDCK) cell monolayers enable comparisons of experiments, simulations and theories  \cite{Reffay2011,Serra-Picamal2012,Harris2012,Doxzen2013,Cochet2013,Albert2016}; 2D images are easier to obtain and analyze than 3D ones, \modif{especially} to extract physical quantities such as cell velocity, density, shape and deformation \cite{Harris2012,Tambe2011}.

When monolayers are grown on a substrate,  the latter acts
as a source of external friction on cells \cite{Trepat2009,Angelini2010,Saez2010,Serra-Picamal2012}.
If it is deformable (made of soft gel or covered with pillars), it acts as a mechanical sensor
for traction force microscopy  to quantify forces exerted by cells on the substrate, which are the opposite of forces exerted by the substrate on the cells \cite{Dembo1996a,Schwarz2002a,Plotnikov2014}.
Beside these external forces, mechanical stresses within the monolayer arise from cell-level processes which include: cell volume change \cite{Zehnder2015} and division \cite{Puliafito2012}; competition between the adhesion to the substrate, the intercellular adhesion and the cell contractility \cite{Hannezo20143Dsheets};
cryptic lamellipodia extending from one cell below its neighbours  \cite{Farooqui2005}.

The emergence of large-scale polarized movements within epithelial cell monolayers largely depends on mechanical factors and external geometrical constraints  
\cite{Angelini2010,Doxzen2013,Tambe2011,Ng2012}.
Loza et al. (using human breast epithelial cells) showed that cell density and contractility control transitions in collective shape, and could predict {\it in vivo} collective migration in a developing fruit fly epithelium \cite{Loza2016}. Microfluidic channel experiments  have shown that the flow velocity of the front can be decomposed into a constant term of directed cell migration superimposed with a diffusion-like contribution that increases with density gradient \cite{Marel2014}.
In the context of a cell monolayer collectively spreading and invading a free space,  highly motile leader cells can appear \cite{Ladoux2016} and locally guide
 small organized cohorts of cells  \cite{Reffay2011}.
The cell velocity decreases with the distance to the moving front \cite{Serra-Picamal2012} while both the cell density   and the  stress   increase with the distance to the moving front \cite{Trepat2009}. 
 Bulk cellular motions also display large-scale coordinated movements of cell clusters that can be seen by the emergence of typically 200~$\mu$m correlation length for the velocity field and large-scale polarization  \cite{Poujade2007,Vedula2012}.

Serra-Picamal \textit{et al.}, by confining cells on a strip then releasing the confinement, observed  two periods of a mechanical wave, propagating backwards from each front, made visible by  oscillations of the cell velocity and its gradient, and suggesting how  stress mediates collective motion  \cite{Serra-Picamal2012}. 
Mechanical force propagation has been reported during the collision of two epithelial cell layers to explain the formation of tissue boundaries \cite{RodriguezFranco2017}.
Similar wound healing experiments displayed a wave of coordinated migration, in which clusters of coordinated moving cells were formed away from the wound and disintegrated near the advancing front; this wave could be amplified by the hepatocyte growth factor~/ scatter  factor
 \cite{Zaritsky2014}. Confluent epithelial cells confined within circular domains exhibit collective low-frequency radial displacement modes as well as stochastic global rotation reversals \cite{Deforet2014,Notbohm2016}.
While oscillations at smaller scales  are common in embryogenesis  (cell size and minute period \cite{Solon2009,Martin2010,Maitre2015,Maitre2016}) or myxobacteria swarms (a few cell sizes, 1 to 100 minutes period \cite{Wu2009}), here in confluent monolayers the oscillation scale is that of a tissue size and of hours, reminiscent of somitogenesis (for review of models, see Ref.~\cite{Oates2009}). 

Even though the appearance of cell coordination and waves in  collective migration experiments is crucial to understand development and associated pathologies, it remains poorly \modif{documented}. 
Migration and division contributions to the front velocity are entangled. Moreover,  cell number is constantly increasing due to cell division, which leads to jamming and slowing of the migration. This usually limits the experiment duration
to a few hours. The experimental uncertainty limits the possibilities of quantitative comparisons with models. The process which determines the velocity direction and amplitude of a cell far from the migrating front is not fully understood. In particular, it is still not clear how cell migration is sensitive to the distance to the migrating front and how cells maintain their polarity far from the migrating front. 
To improve our understanding, distinguish between the models, and constrain their parameters, varied and controlled experimental data are required.

Here, we significantly improve  experimental reproducibility and signal to noise ratio, and provide a precise analysis (Section \ref{sec:materials}). We observe a coherent collective cell migration over several millimeters,  quantify average cell velocity profile and waves that develop on top of it, and identify the roles of density and lamellipodia  (Section \ref{sec:results}).
Our \modif{minimal} model of   strain and polarity coupling suggests an interpretation of these experimental observations
 (Section \ref{sec:model}).
Finally, by discussing and comparing the experiments and the model, we quantitatively confirm several preliminary results found in the literature,  and add new results and  insights regarding the role of mechanics in collective cell migration  (Section \ref{sec:discussion}).

\section{Materials and methods}
\label{sec:materials}

\modif{Section \ref{sec:experiments} explains how}, by inhibiting cell division, we see a decrease in cell density due to migration.
We observe a steady collective migration over a day or more, without reaching jamming densities. 
We focus on such long-distance migration and the strip length is adapted accordingly.
Strips are narrow to prevent front shape instabilities, and the cell flow is essentially \modif{one}-dimensional
(Fig.~\ref{fig:monolayer_contour_velocity_density}, Fig.~S1 and
Movies~S1-6 in the Supporting Material). 
\modif{Section \ref{sec:analysis} explains how w}e first average out the velocity field over several hours, to characterize the mean cell velocity profile in the monolayer bulk. We then  quantify the fluctuations around this average with an unprecedented signal-to-noise ratio using wavelet analysis.

\subsection{Experiments}
\label{sec:experiments}


The micropattern of fibronectin is printed according to the following standard soft lithography technique, robust to small changes in the procedure.
Patterned PDMS stamps prepared from silanized wafers are incubated for 45~min at 37$^\circ$C  
or 60~min at room temperature  
 with a solution of 50~$\mu$g/mL of fibronectin (Sigma) and 25~$\mu$g/mL of Cy3 conjugated fibronectin.  A thin layer of \modif{rigid} PDMS (10 \% of reticulating agent) is spin coated on a 35 mm plastic Petri dish and cured for 2~h  at 80$^\circ$C  
 or overnight at 65$^\circ$C. The Petri dish is exposed to UV for approximately 20 min in order to activate the PDMS surface. After incubation, stamps are dried and pressed on the UV activated PDMS surface in order to transfer fibronectin. A 2 \% Pluronic F-127 (Sigma) solution is added to the Petri dish to chemically block the regions outside of the fibronectin pattern for 1 hour at room temperature.
The Pluronic solution is removed after 1 hour and the Petri dish is rinsed 3  
to 6 
times with a PBS solution.


We use the same MDCK strain as in Ref.  \cite{Vedula2012}.
A stable cell line was created using Histone GFP \cite{Fukunaga2005} using the DreamFect Gold Transfection Reagent, Oz Biosciences.

A batch has 3 to 6 strips, with  the same initial MDCK cell density,  lengths up to 4~mm. Strip widths range from 200~$\mu$m to 1~mm (at least equal to the typical 200~$\mu$m correlation length for the velocity field \cite{Poujade2007,Vedula2012}) and do not affect the results presented here. 
Different batches correspond to different initial cell densities, tested at least twice each.

Suspended  cells are deposited and allowed to attach for one  
to a few  
 hours. Non-attached cells are rinsed, while attached cells grow and divide until full confluence.
The confining PDMS block  is removed. Some cells might detach, so the monolayer is rinsed again and left for a few hours. The monolayer starts to migrate along the whole accessible strip, expanding towards the empty surface where cells adhere to fibronectin, and not towards outside regions chemically blocked using Pluronic.

 \modif{To vary the initial cell density, we  vary the amount of deposited cells and/or  the time they are left to proliferate; we always begin with confluent monolayers.}
\modif{We do not measure cell volumes; we expect they are all similar at the time of deposition and that the main contribution to their variation is that of cell cycles.}
\modif{When the imaging begins  and the density is measured, the monolayer has already migrated for a few hours (see below), so the initial density varies spatially from the reservoir to the front (for details, see caption of Fig.~\ref{fig:profiles}A,B).}

In order to decrease the division rate, 8~$\mu$L of a 0.5~mg/mL mitomycin~C solution \modif{(aliquoted, stored at $-20^\circ$C, and used within a day after thawing)} is added to 1~mL cell culture medium, and cells are incubated at 37$^\circ$C for 1~h \cite{Carretero2008,Deforet2014}. 
They are then abundantly rinsed with fresh 37$^\circ$C medium to prevent the toxicity effects reported for 12~h exposure to mitomycin \cite{Poujade2007}. 
After 3~h the division rate is less than a fifth of the initial one (Fig.~S2 in the Supporting Material), and the rate of extrusions also strongly decreases. 
Control experiments \modif{are performed} in standard conditions, with proliferating cells (no mitomycin C added).

To test the role of lamellipodia, we prepare a 100~$\mu$M solution in DMSO of CK666, namely 2-Fluoro-N-[2-(2-methyl-1H-indol-3-yl)ethyl]benzamide, a selective inhibitor of actin assembly mediated by actin-related protein Arp2/3 (IC50 = 17~$\mu$M) \cite{Nolen2009,Wu2012,Vitriol2015}. Aliquots are stored at $-20^\circ$C and used within two weeks of preparation. The solution is added to the cells after $\sim$1~day of migration and is not rinsed. Lamellipodia (both cryptic and front ones) are no longer detectable (Movie~S7 in the Supporting Material).
 

Two hours after having added the mitomycin, we take the first image of the movie and define it as $t=0$. Live imaging of monolayers is performed  in the Nikon BioStation IM, a compact cell incubator and monitoring system, with an air  objective (CFI Plan Fluor 10X, Nikon). Phase-contrast and fluorescent imaging are used to observe respectively cell contours and cell nuclei.
The interframe time interval is 5~min for 1~mm wide strips  
and 6~min for 200~$\mu$m wide strips.

Dead or extruded cells appear as bright spots which can be removed by manual image intensity thresholding. 
The contrast is adjusted separately on each color channel, and a blur with 2 pixel radius removes sharp intensity fluctuations.
To obtain the whole view of the confined monolayer, up to 6  (for two 200~$\mu$m wide strips) or 20  (for 1~mm wide strips) microscope fields of view are merged using the Grid/Collection Stitching Plugin \cite{Preibisch2009} implemented in ImageJ.  We use the "unknown position" option for the first time frame to calculate automatically the overlap between images, which we use for all frames since images are stable.

\subsection{Data analysis}

\label{sec:analysis}


We measure the two-dimensional velocity field  $\vec{v}(x,y,t)$  (Fig.~\ref{fig:monolayer_contour_velocity_density}B and Fig.~S1A,C)  using Particle image velocimetry \cite{Vig2016}.
We use the open source toolbox MATPIV \textit{matpiv.m} \cite{matpivref} of Matlab (The MathWorks, Inc., Natick, Massachusetts, United States), with the "singlepass" option, square box of side 32 pixels (20~$\mu$m) for  200~$\mu$m wide strips and 128 pixels (80~$\mu$m) for 1~mm wide strips, and box overlap is 50~\% or 75~\% for both widths.

The Particle image velocimetry method, option "single", interrogation box size of 128 pixels, yields qualitatively identical results, and is quantitatively around 10\% larger, when compared either with "multin" option, windowsize-vector  [128~128~;~64~64] or with Kanade-Lucas-Tomasi (KLT) feature matching algorithm, pyramid parameter 2, successive interrogation box sizes of 128 and 64 pixels.
 
We do not detect any statistically significant dependence of $\vec{v}$ with $y$, even near the lateral sides of the strip. 
The $y$ component of $\vec{v}(x,y,t)$ 
  is lower at higher positions $x$ where the average velocity is higher (Fig.~S1C), indicating a more directed movement; we do not consider this component in what follows.
  
The component of $\vec{v}(x,y,t)$ along the $x$ axis, i.e. along the long axis of the strip, averaged over $y$, is  
 the one-dimensional velocity field ${V}(x,t)$, which we study here. 
 \modif{This first step, projecting $\vec{v}$ on $x$ and averaging it over $y$, is already enough to make visible the main features of the velocity field: velocity gradient along $x$ and propagating waves.}
 
 \modif{To improve the visualisation, and enable for a qualitative analysis, we} plot the space-time diagram or "kymograph"  of  $V(x,t)$\modif{. The next step consists in filtering it. We first remove small scale} noise using  a Gaussian blur of standard deviation 15~min and 30~$\mu$m (and a sliding window which is three times larger). We then separate scales, and decompose  \modif{this denoised $V$ into large-scale $\bar{V}$ and middle-scale $V-\bar{V}$}, using a Gaussian filter of standard deviation 50~min and 100~$\mu$m (again, with a sliding window which is three times larger).  

For large scale profiles $\bar{V}$, discrete measurements used for graphs are performed with an average over $176$~$\mu$m wide \modif{space boxes}  in the distance \modif{$d$} to the moving front (this  coordinate \modif{$d$}  is oriented from the migrating front toward the cell reservoir, as opposed to the coordinate $x$)\modif{; and an average on time $t$ over 180~min.} We entirely exclude the first \modif{box}, where statistics are noisy due to the front. 
Only significant data points are plotted, i.e. points with enough pixels in the 176~$\mu$m $\times$ 180~min \modif{space-time} \modif{bin} (at least 150 pixels, out of a maximum of 612) and where the signal value is larger than its SD. For the velocity gradient, a finite difference gradient is used and the resulting very small scale noise is removed with a 3-pixels wide linear filter. 
 
Similarly, by using histone-GFP (Fig.~S1B), we identify cell nuclei. Our measurements are based on local maxima detection, independently of the maximal 
intensity value, and thus are not sensitive to possible variations in the GFP signal \modif{intensity}. The nuclei density can vary by a factor 10 within a given image. We first use a low blur radius, optimised for the highest nucleus density on the image. It yields a good detection for high density regions using \textit{FastPeakFind.m} but gives false positives (more than one local maxima per nucleus) for low density regions. When the distance between two maxima is smaller than a critical value (equal to a third of the local average distance between nuclei), we remove the less intense one. According to manual checks on high, middle and low density regions, the precision is better than 5~\%.

We have checked that tracking the cell nuclei yields more fluctuations than PIV for $\vec{v}(x,y,t)$ measurements,  due to intra-cellular movements of nuclei (Movies~S3-6), but  \modif{once projected and averaged it} yields same results as PIV for  ${V}(x,t)$. We use this cell nuclei detection to plot the cell density $\rho(x,t)$,  using boxes $(x,y)$  of 20~$\mu$m $\times$ 20~$\mu$m, then an
average over $y$. Using the same filters as for $V$, we remove noise, decompose $\rho$ into $\bar{\rho}$ and $\rho-\bar{\rho}$, and define discrete measurements of  \modif{density  with an average on time $t$ over 180~min and on space over $176$~$\mu$m wide boxes  in the distance $d$ to the moving front. We entirely exclude the first box, where statistics are noisy due to the front. In Figs.~\ref{fig:diagphase}C, S3B, the average near the front is taken on the three boxes ($3 \times 176 = 528$~$\mu$m) next to the front one.}


\modif{To perform quantitative analyses of the kymographs, we use wavelets as measurement tools (rather than as filtering tools). They  extract from a signal its wavelengths and time frequencies, like Fourier Transform does, but in addition wavelets can determine the space and time variations of these quantities. We use  a custom-made software for wavelet transform profilometry (WTP)}, a method inspired by 3D-fringe projection profilometry~\cite{gdeisat_spatial_2006, gdeisat_spatial_2009}. 
It involves a one-dimensional continuous wavelet transform with a phase estimation approach. It is reliable, easy to implement, and robust to noise.  
We choose a Morlet wavelet and the wavelet transform is computed using a FFT algorithm (which  is equivalent to an analytic Morlet Wavelet) with a Matlab script \cite{continuous_wavelet}.

For each kymograph line (i.e. for each fixed position $x_i$), the signal  wavelet transform is computed at various time scales, in an observation window \modif{of} 80 to 400~min, chosen in order to cover the full range of characteristic times of the observed oscillations (we checked that this choice does not affect the results presented here). The wavelet transform returns a matrix  of complex coefficients $A(x_i,t,s)$, defined as continuous wavelet coefficients where $s$ represents the test times scales. Each coefficient provides a local measurement of the similarity between the signal and the wavelet at a scale $s$. For each point $(x_i,t_j)$ of a given line $i$ in the kymograph, only the coefficient $A_m(x_i,t_j)$ having the largest modulus with respect to the scale $s$ is kept. 

The argument of $A_m(x_i,t_j)$ provides the wrapped fringe phase $\phi_w(x_i,t_j)$. The phase $\phi_w(x_i,t_j)$ is unwrapped along time, and the local \modif{angular frequency $\omega$ ($2\pi$ times the frequency)} is deduced by differentiation with respect to time $t$ \modif{according to the sign convention $\omega=\partial \phi_V / \partial t$}. Independently, the phase $\phi_w(x_i,t_j)$ is unwrapped along space, and the local  wave number $k$ is deduced by differentiation with respect to space $x$ \modif{according to the sign convention $k=-\partial \phi_V / \partial x$}.

\section{Results}
\label{sec:results}

\modif{Section \ref{sec:velavgtime} reports} and quantif\modif{ies} coherent collective cell migration over several millimeters. \modif{A priori, one could expect the cell velocity to depend on density, density gradient and distance to front (hence on the monolayer history). In fact, t}he mean cell velocity profile in the monolayer bulk depends explicitly only on the cell density, irrespectively of the distance to the migrating front\modif{. It} is very sensitive to \modif{proliferation and to} lamellipodia inhibition. 
\modif{Section \ref{sec:velwaves} reports that, on top of the average velocity and density profiles}, backwards propagating waves in density and velocity exist, and they have an opposite phase.
We measure the local characteristics of the velocity waves and their variation in space. Their velocity decrease\modif{s} with cell density. 
Inhibiting lamellipodia formation \modif{damps the waves and decreases their frequency} or even leads to their suppression. 

\subsection{Large scale profiles of velocity and density}
\label{sec:velavgtime}
 
\begin{figure}[h!]
\noindent (A) \hfill     (B) \hfill  ~\\
\showfigures{\includegraphics[width=1\columnwidth]{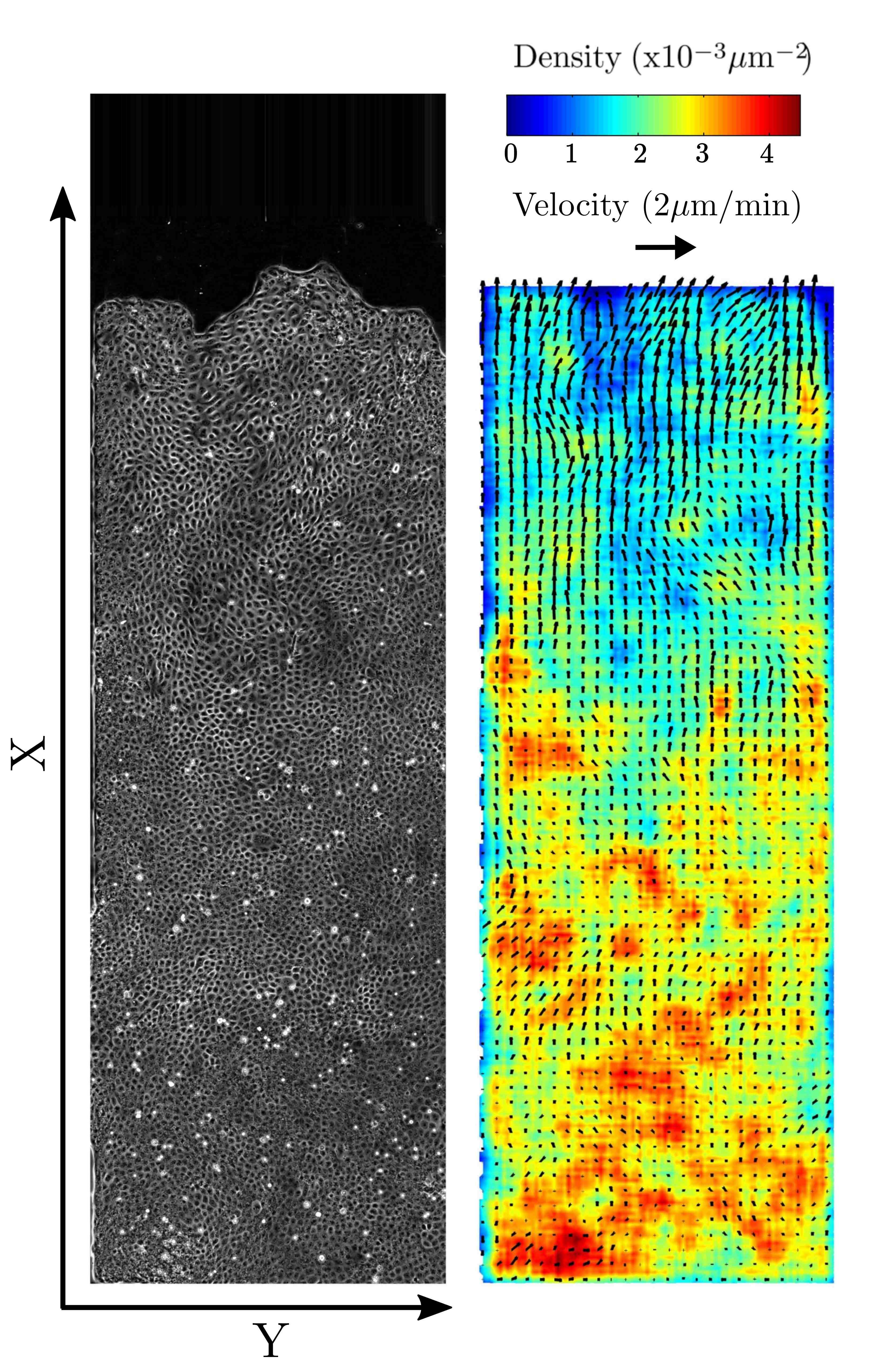}}
\caption{Cell migration.
{\it (A)}
A monolayer of MDCK cells, initially confined, is released. It expands (Movies~S1-3) along the adhesive strip towards empty space (direction of increasing $x$). 
Mitomycin~C is added to inhibit divisions.
Phase-contrast image of  cell contours,  taken at $t=11$~h 30~min (i.e. after $\sim$16~h 30~min of migration).
Strip total length 4~mm (most of it is visible here), width 1~mm.  
{\it (B)} Corresponding 2D fields of cell velocity and density.
Scale arrow: 2~$\mu$m/min.
\label{fig:monolayer_contour_velocity_density}
}
\end{figure}

By averaging over the direction $y$ perpendicular to the stripes, we determine the one-dimensional cell velocity field ${V}(x,t)$ and cell density field $\rho(x,t)$. We first investigate their overall profiles $\bar{V}$, $ \bar{\rho}$, obtained by large scale sliding average.
Due to the spreading of cells, variations of  velocity and density are visible (Fig.~\ref{fig:monolayer_contour_velocity_density}B). 
Far from the front (bottom of Fig.~\ref{fig:monolayer_contour_velocity_density}B), the density is still close to its initial value and the velocity is still zero, while close to the front (top of Fig.~\ref{fig:monolayer_contour_velocity_density}B) the density has decreased and the velocity increased. 
Note that far from the front, velocities are occasionally negative.
We introduce $R = \left( \pi \rho \right)^{-1/2}$, then $R_\mathrm{mean} = \left( \pi \bar{\rho} \right)^{-1/2}$, interpreted as a  mean effective cell radius.
Its typical range of variation is 8 to 15~$\mu$m; for comparison, note that 4.8~$\mu$m corresponds  to the nuclei being almost close packed, 
while 17~$\mu$m is the radius of a front cell at the limit of detaching from the monolayer.

\begin{figure}[h!]
\noindent (A) \hfill    (B) \hfill  ~\\
\showfigures{\includegraphics[width=0.5\columnwidth]{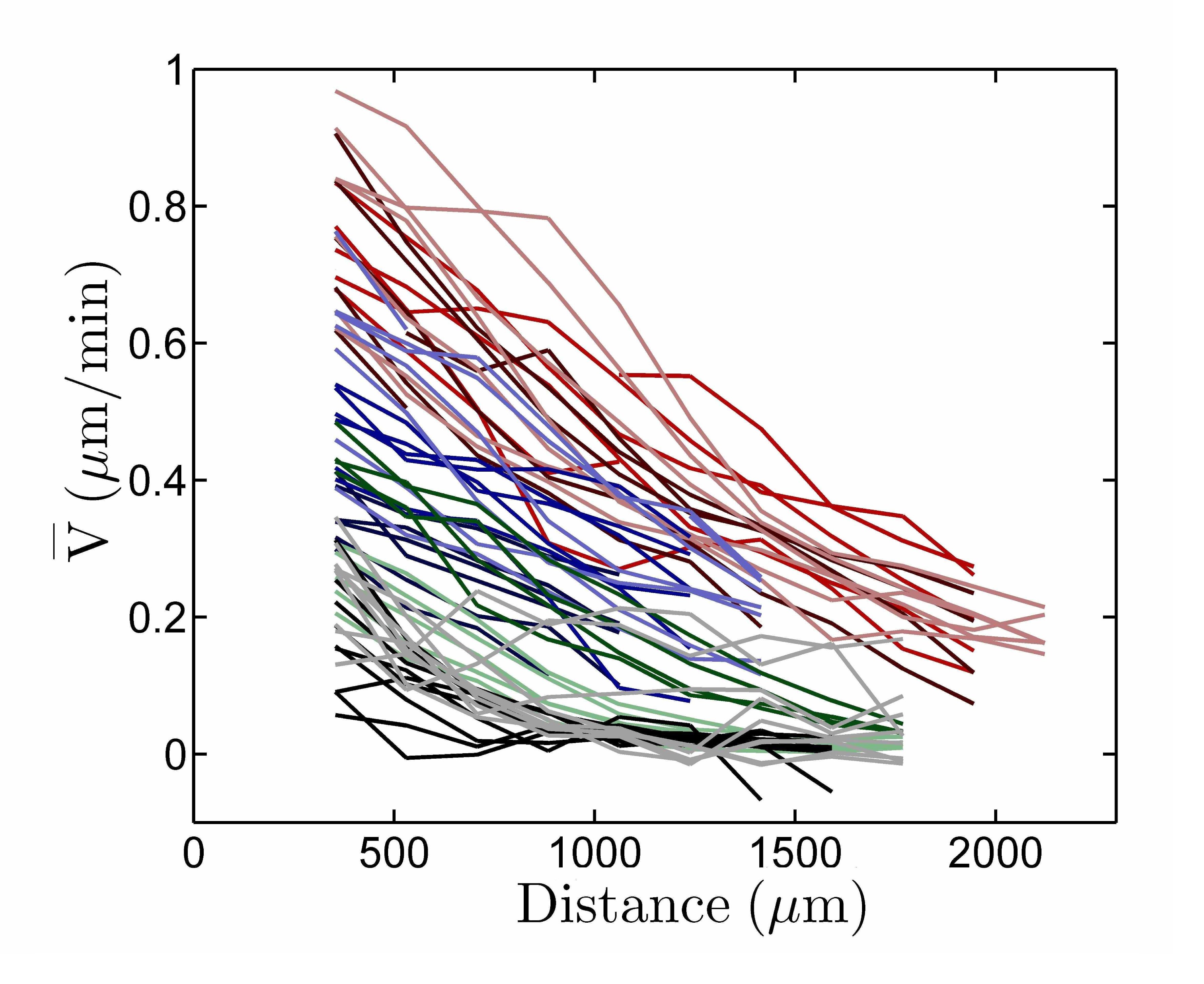}}
\showfigures{\includegraphics[width=0.5\columnwidth]{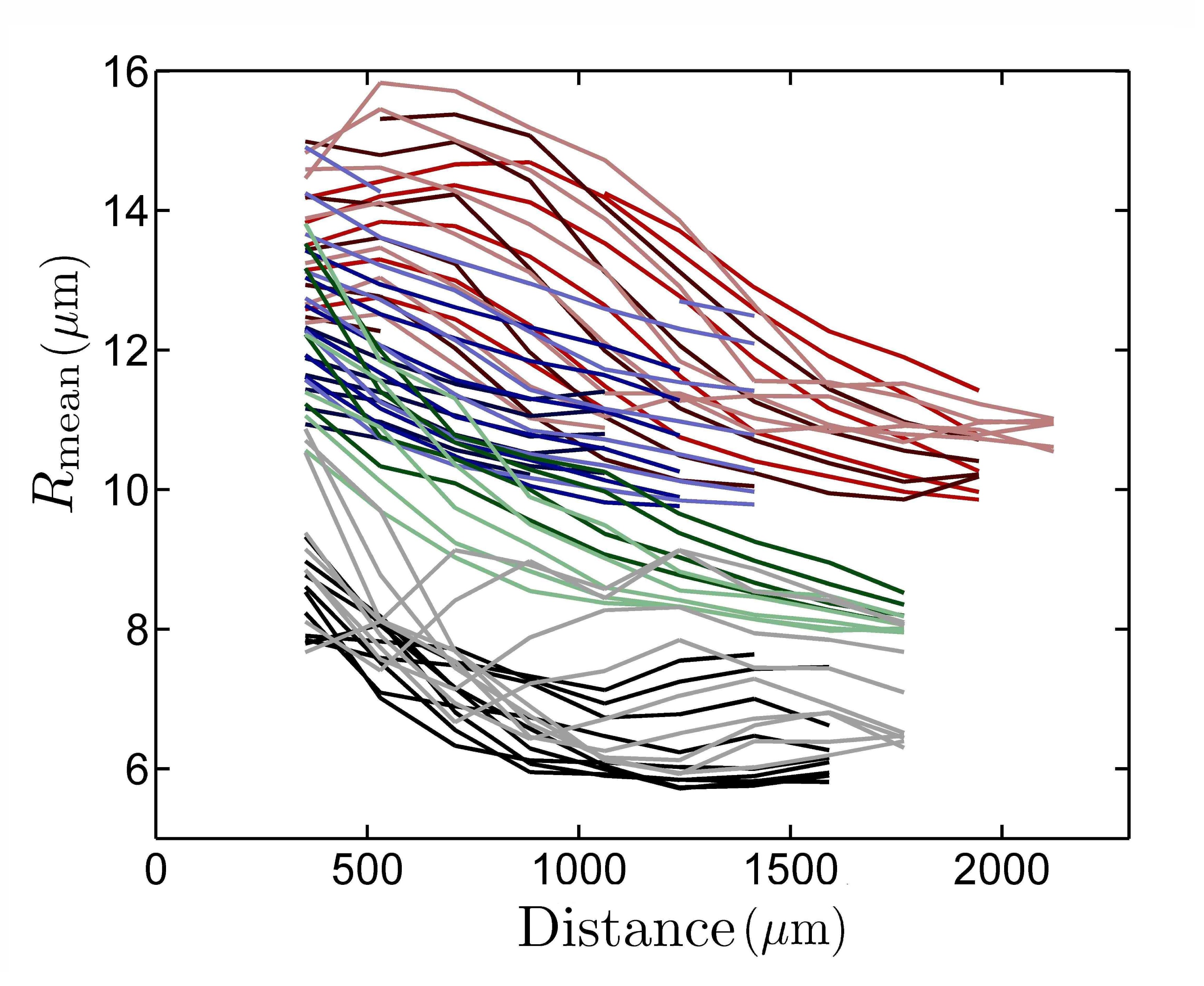}}
\\
\noindent (C) \hfill    (D) \hfill  ~\\
\showfigures{\includegraphics[width=0.5\columnwidth]{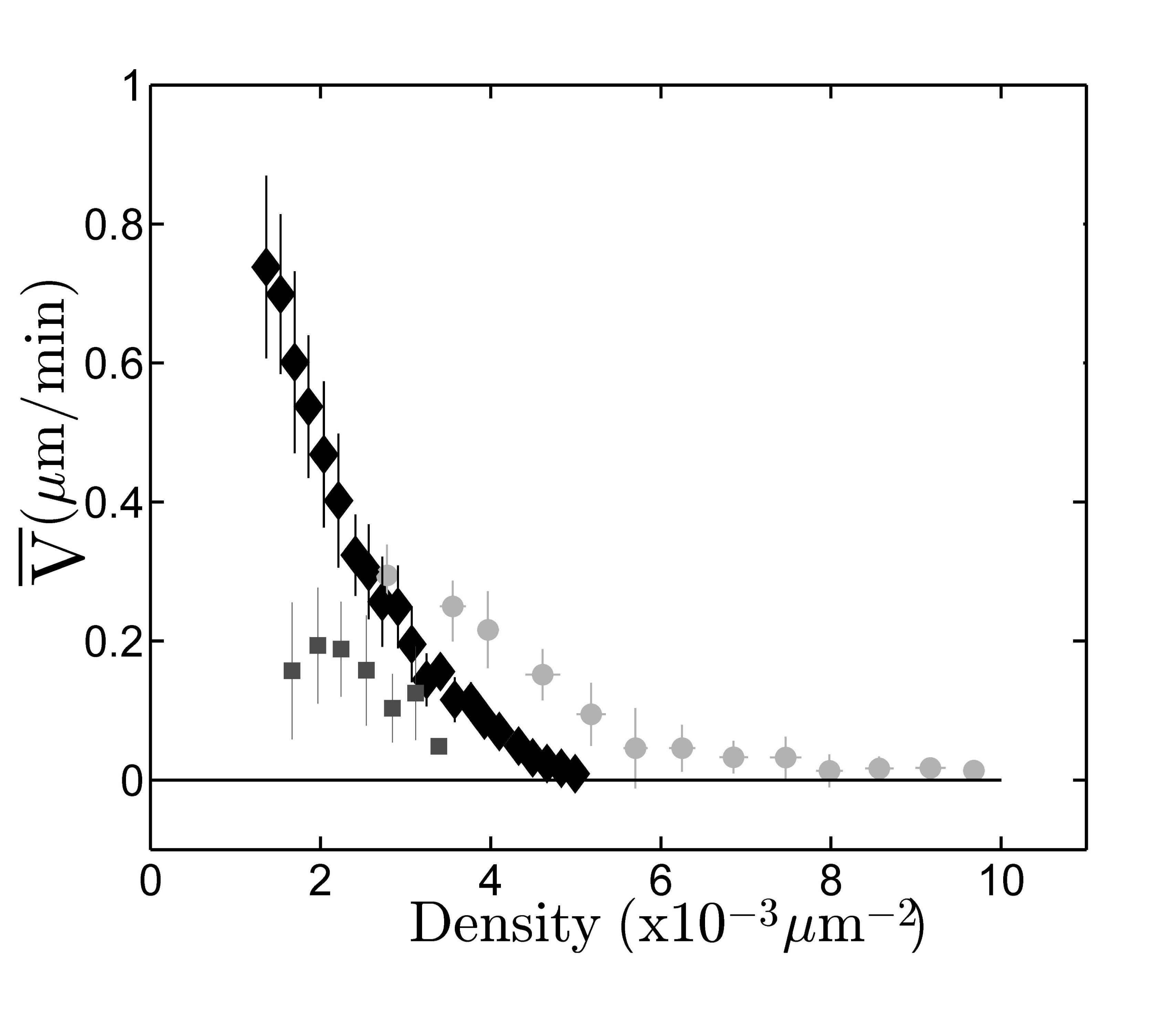}}
\showfigures{\includegraphics[width=0.5\columnwidth]{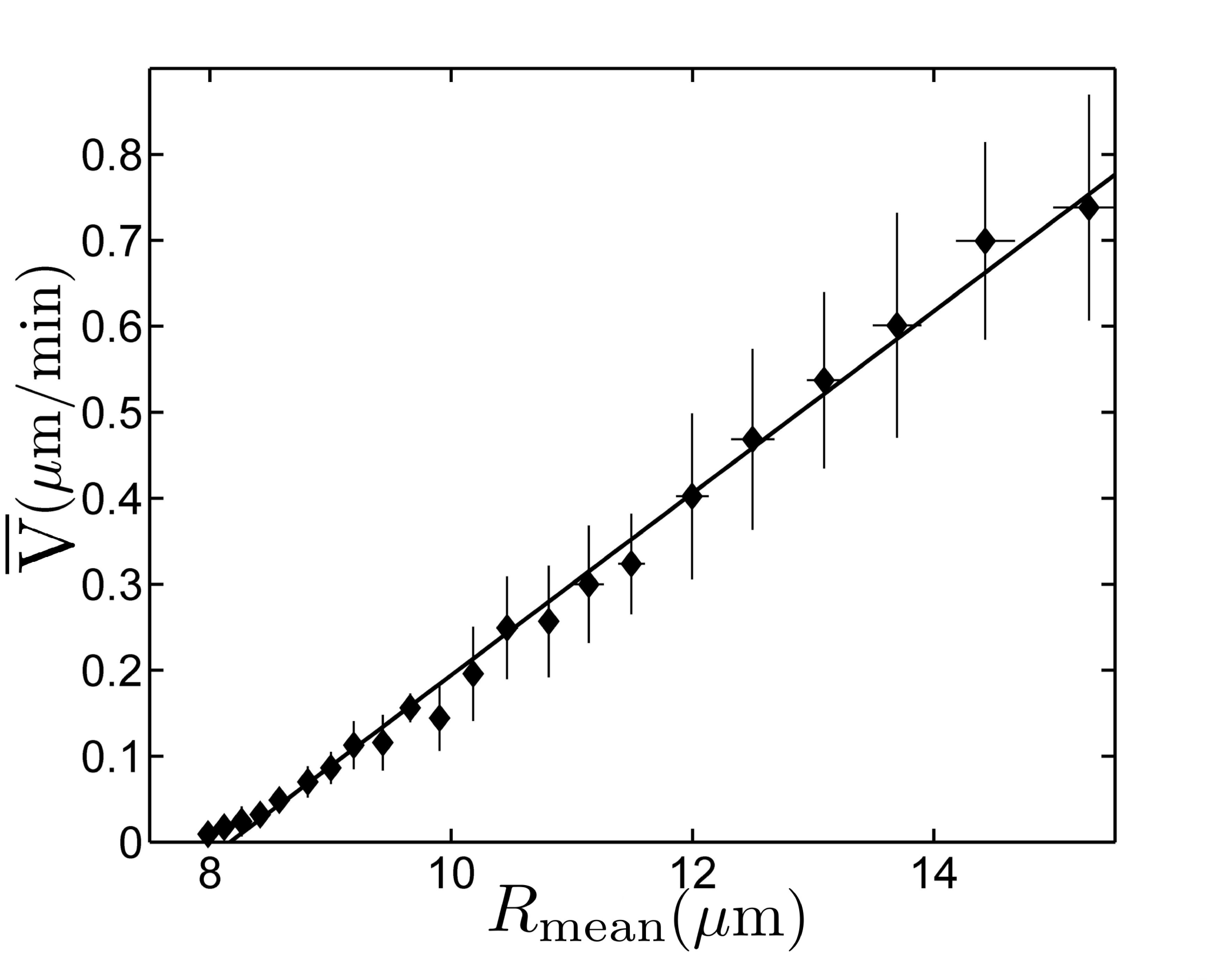}}
\caption{ \modif{Cell velocity and density profiles.}
{\it (A, B)} Large scale profiles.
Large-scale average of  (A) cell velocity $\bar{V}$,   and   (B) mean effective cell radius $R_\mathrm{mean} = \left( \pi \bar{\rho} \right)^{-1/2}$, 
plotted vs distance \modif{$d$}  to \modif{moving} front (oriented from the front toward the cell reservoir).
Each color marks a different batch,  with initial density \modif{(from reservoir to front):  red, 10 to 12.5 $\mu$m; blue, 10 to 11 $\mu$m; green, 8 to 10.5 $\mu$m.} 
For a given color (i.e. batch), each shade marks a different strip. 
For a given shade (i.e. strip), each data point is the average in a 176~$\mu$m $\times$ 180~min \modif{bin}.
Grey \modif{and black curves} are the same, for control experiments without mitomycin~C\modif{, with initial density (from reservoir to front):  grey, 8 to 10 $\mu$m; black, 7.7 to 7.9 $\mu$m.} 
\modif{{\it (C,D)} Cell velocity - density correlation.}
Same data \modif{as in (A,B)},  with mitomycin, binned and plotted as $\bar{V}$  vs cell density $\bar{\rho}$ (black diamonds) (C) or vs $R_\mathrm{mean}$ (D), with a linear fit $\bar{V} = 0.106 \; R_\mathrm{mean} - 0.864$ (R=0.9931); $N=8$~strips. 
Light grey circles: control experiments without mitomycin~C; $N=3$~strips. Grey squares: experiment with CK666  \modif{drug} to inhibit lamellipodia; $N=5$~strips. Horizontal and vertical bars: standard deviation (SD) within each \modif{bin}. 
\label{fig:profiles}
}
\end{figure}

With respect to the distance \modif{$d$} to the moving front, different experimental batches display 
 $\bar{V}(\modif{d})$ profiles  which are qualitatively similar and quantitatively different (Fig.~\ref{fig:profiles}A), and $R_\mathrm{mean} (\modif{d})$ profiles too (Fig.~\ref{fig:profiles}B). When eliminating the space variable \modif{$d$}, points coming from different batches fall on the same curve: 
 $\bar{V}$ has a strong, negative correlation with $\bar{\rho}$ \cite{Doxzen2013} (Fig.~\ref{fig:profiles}C; Fig.~S3A in the Supporting Material), decreasing  from 0.8~$\mu$m/min to 0 for $\bar{\rho}$ ranging from 1.8 to 5~10$^{-3}$~$\mu$m$^{-2}$.
In fact, $\bar{V}$  increases linearly  \modif{(with a non-zero intercept) with the mean effective cell radius, $\bar{V} = 0.106 \; R_\mathrm{mean} - 0.864$} (Fig.~\ref{fig:profiles}D). 
This relation does not depend on the distance to the front, and is unaffected when the sliding window size  in time is doubled.
 
We also performed experiments under standard conditions (i.e. without mitomycin~C, Fig.~S4 in the Supporting Material). 
While the overall relation between $\bar{\rho}$ and $\bar{V}$  seems qualitatively unaffected, dividing cells have a significantly larger $\bar{V}$ at given $\bar{\rho}$, and a larger  \modif{"arrest density" defined as the intercept of the velocity graph with the density axis}: $\bar{V}$  decreases  from 0.3~$\mu$m/min to 0 for $\bar{\rho}$ ranging from  3  to 10~10$^{-3}$~$\mu$m$^{-2}$ (Fig.~\ref{fig:profiles}C; Fig.~S3A).

\modif{Inhibiting lamellipodia formation drastically decreases the monolayer average velocity (Figs.~\ref{fig:profiles}C, \ref{fig:kymo_drugs})}.

\subsection{Propagating waves}
\label{sec:velwaves}
 
\begin{figure}[h!]
\begin{center}
\noindent (A) \hfill   ~\\
\vspace{-6mm}
\showfigures{\includegraphics[width=0.7\columnwidth]{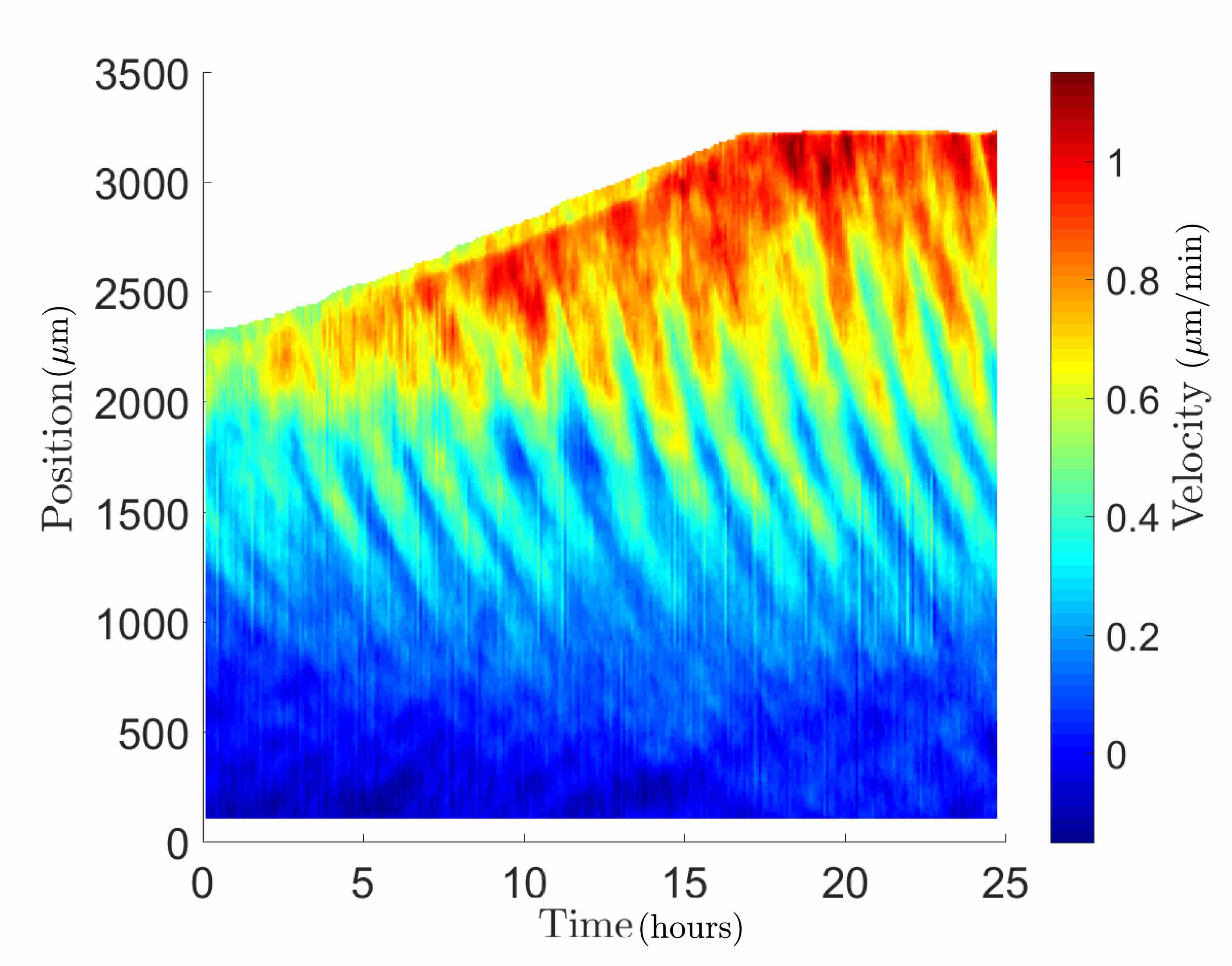}} 
\\
\vspace{-6mm}
\noindent (B)  \hfill  ~\\
\showfigures{\includegraphics[width=0.7\columnwidth]{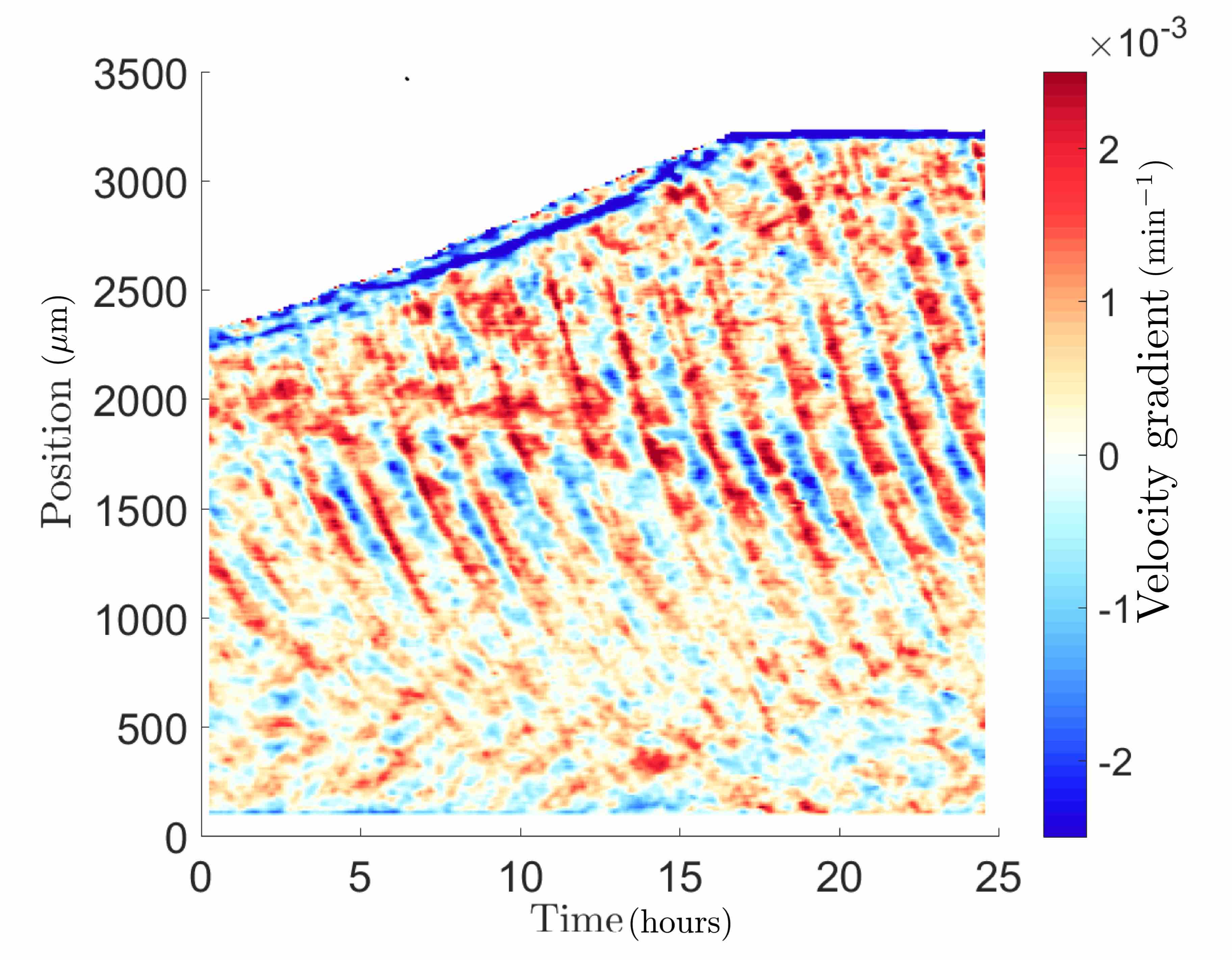}} 
\\
\vspace{-6mm}
\noindent (C) \hfill  ~\\
\showfigures{\includegraphics[width=0.7\columnwidth]{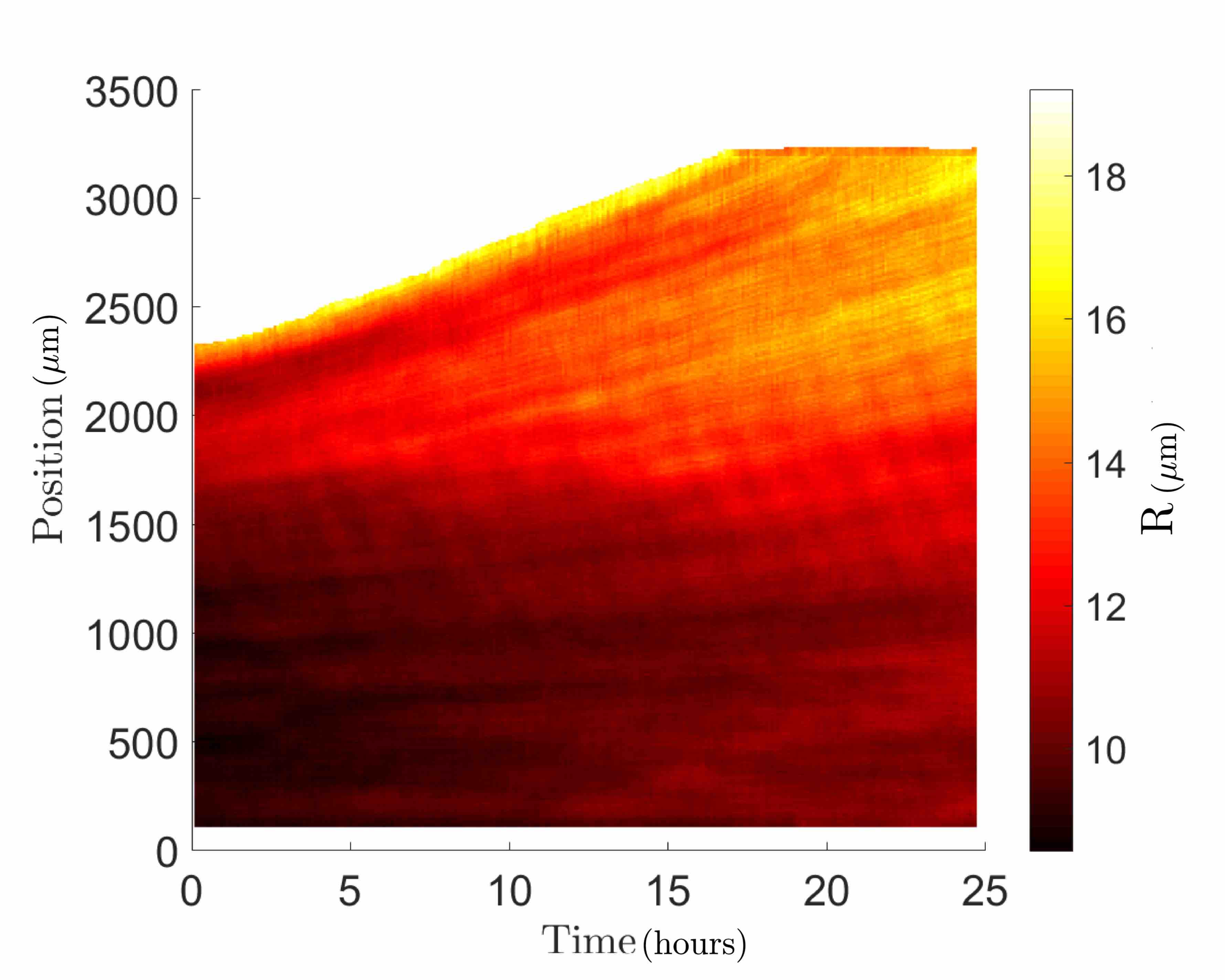}} 
\end{center}
\caption{ 
Propagating waves.
Space-time diagram ("kymograph") of  {\it (A)} cell velocity $V$, {\it (B)} velocity gradient  $\partial \widetilde{V} / \partial x$,  and {\it (C)} effective cell radius  $R$. 
Space $x$ is oriented from cell reservoir (bottom, 0~mm) to   toward the front (top, 3~mm),  time $t$ from left (0~h) to right (25~h), and the top-left region is the bare substrate in front of the monolayer. All 8 strips showed similar results.
\label{fig:kymo_wt}
}
\end{figure}

We now turn to \modif{middle-}scale variations.
The cell velocity  $V(x,t)$ displays waves: cells slow down and accelerate while  waves propagate  from the front backwards in the $-x$ direction (Movies~S1-6).
In the moving frame of average cell velocity, these waves would appear as a periodic velocity reversal. They are visible quantitatively \modif{even on the raw} kymograph  (Fig. \ref{fig:kymo_wt}A, Fig.~S5A in the Supporting Material), and more clearly on the velocity \modif{middle-}scale variations $\widetilde{V} = V - \bar{V}$  (Fig.~S6A in the Supporting Material) as well as on the velocity gradient 
(Fig. \ref{fig:kymo_wt}B).  The waves are reproducibly observed  near the front, with a good signal-to-noise ratio
 over more than ten periods for a  whole observation duration of $\sim$~20~h.  We do not detect any particular effect of the strip width on waves (compare Figs.~S4B, S5A).

First measurements are manual. They indicate that the waves have a period around two hours, their wavelength is around one millimeter. 
Their amplitude decreases with the distance to the front: waves are not apparent \modif{near the reservoir},  where the cell density is as high as 5~$10^{-3}$~$\mu$m$^{-2}$ and the cell velocity vanishes.  
Where they are visible, their amplitude is steady in time, and large: it represents a relative variation in local velocity which typically ranges from  15\% to 30\%  (Fig.~S5B), i.e. up to 60 \% crest-to-crest.
Their velocity (indicated by the slope of the \modif{wave pattern}) is of order of ten micrometers per minute, and with a sign opposed to that of the cell velocity (indicated by the slope of the front position). 
\modif{The wave pattern is}  visibly curved: this evidences that the wave velocity (phase velocity) is larger near the front than in the middle of the monolayer.

Density \modif{middle-}scale variations are dominated by local  heterogeneities, which are signatures of  initial density fluctuations (cells do not significantly mix nor rearrange) over a typical length scale of 200~$\mu$m. 
More precisely, on the kymograph of $R$ or $\widetilde{\rho} = \rho - \bar{\rho}$ these fluctuations  appear as bars which, near the front, are almost parallel to the front;  far from the front, they are closer to horizontal (Fig. \ref{fig:kymo_wt}C, Fig.~S6B); in between, along the line drawn on Fig.~S6C which has a slope of 0.31~$\mu$m/min, we measure $\bar{V}=0.33\pm 0.02$~$\mu$m/min (SD). This proves that 
these fluctuations are advected at local velocity $\bar{V}$ along with the monolayer itself.
In addition, and although they are less visible, it is clearly possible to distinguish waves on the density (Figs. \ref{fig:kymo_wt}C, S5C, S6B), which have the same period as  the velocity waves and are in phase opposition with them (Fig.~S6C,D). They have a small amplitude, with  a relative variation ranging from  1\% to  2 \%  (Fig.~S5D), i.e. up to 4 \% crest-to-crest.

\begin{figure}[h!]
\noindent (A) \hfill    ~\\
\showfigures{\includegraphics[width=1\columnwidth]{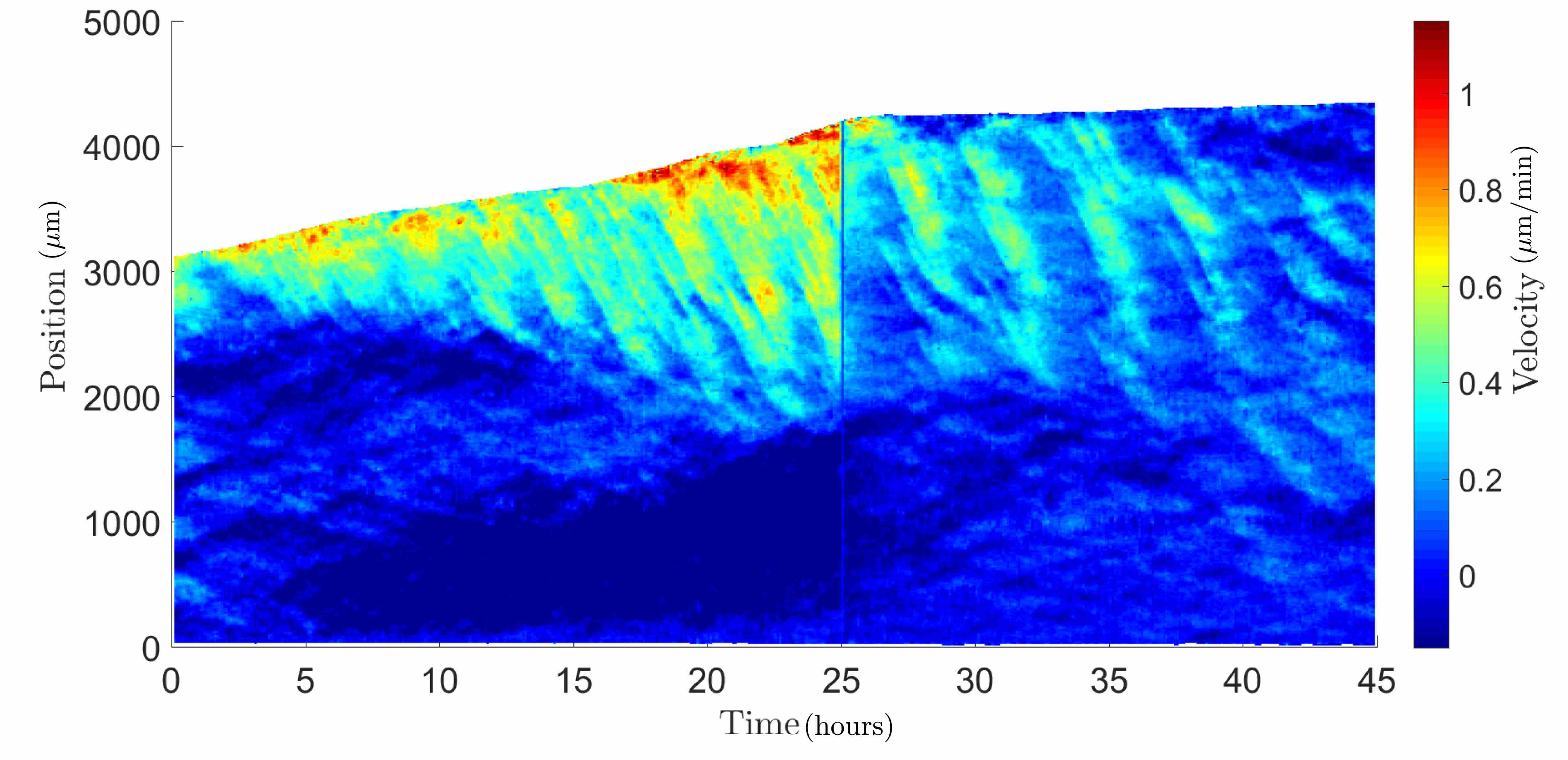}} 
\\
\noindent (B) \hfill  ~\\
\showfigures{\includegraphics[width=1\columnwidth]{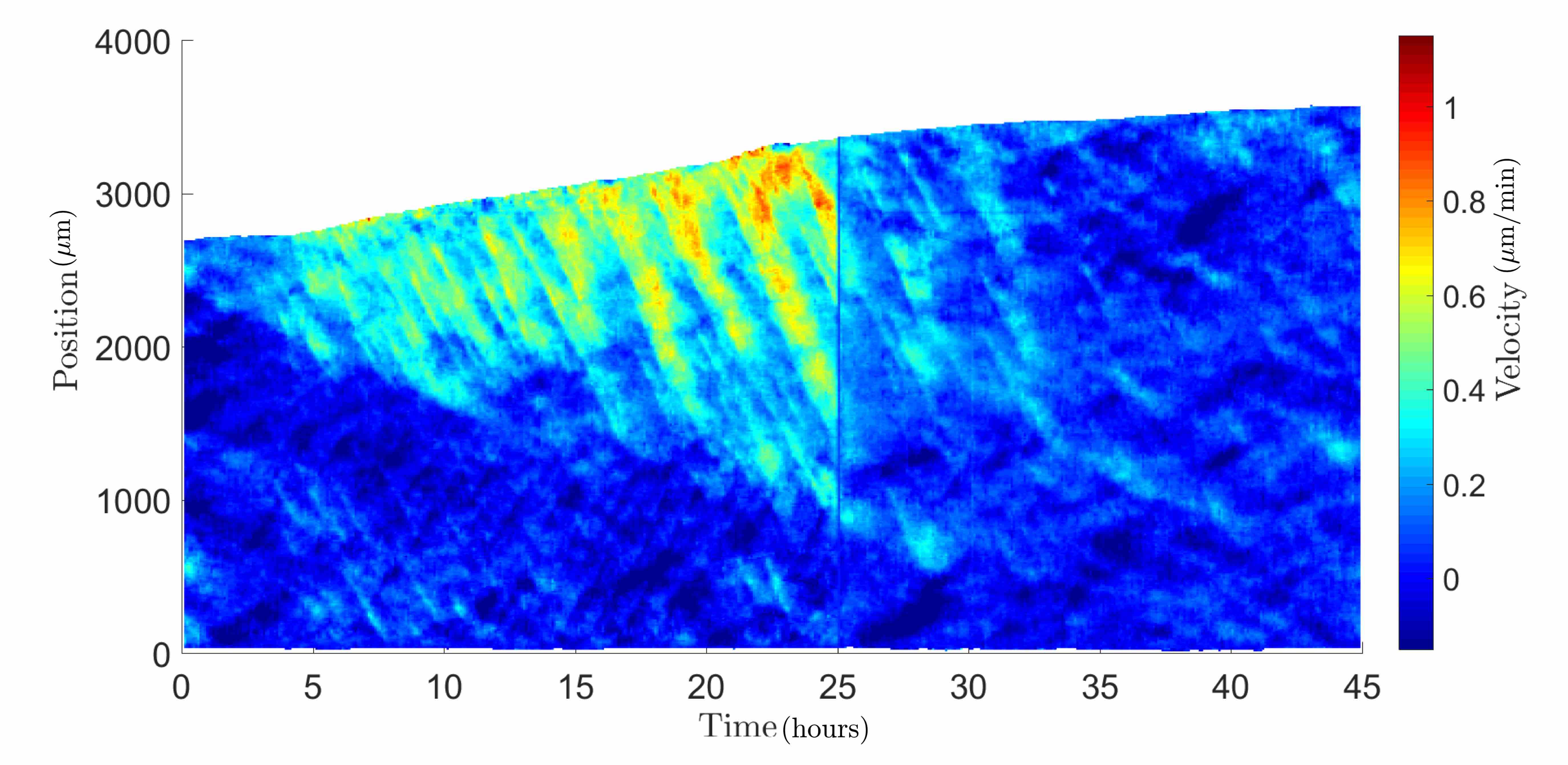}} 
\caption{ 
Propagating waves, \modif{same as Fig.~\ref{fig:kymo_wt},} with application, after one day, of CK666 drug to inhibit lamellipodia formation, resulting in {\it (A)} attenuation of waves, $N=7$~strips; or {\it (B)} in their almost complete suppression, $N=5$~strips.
Space $x$ is oriented from cell reservoir (bottom, 0~mm) to   toward the front (top, 3~mm),  time $t$ from left (0~h) to right (\modif{45}~h), and the top-left region is the bare substrate in front of the monolayer. 
\modif{The time of drug application
(25~h) is visible as a vertical bar, since one image is not recorded.}
\label{fig:kymo_drugs}
}
\end{figure}

Inhibiting lamellipodia formation strongly decreases the amplitude \modif{and frequency} of velocity waves  (Fig. \ref{fig:kymo_drugs}A), sometimes even almost completely suppressing them (Fig. \ref{fig:kymo_drugs}B). The wave velocity, visible as the slope of the \modif{wave patterns}, is not significantly altered  (Fig. \ref{fig:kymo_drugs}A).
 
\label{sec:wavelet}

\modif{The wave pattern is}  visibly curved (Figs.~\ref{fig:kymo_wt}A, S6A): wave characteristics vary with space, and this can be quantified in several regions where the signal-to-noise ratio is sufficient  (Fig.~S5A). 
Using wavelets, we define, distinguish and measure locally $\vert \widetilde{V}\vert $  and $\phi_V$ at each position $x$ and time $t$, as follows.
The  small\modif{er} space and time scales variations \modif{of $ \widetilde{V}$} are encompassed by  \modif{the wave phase} $\phi_V$  (Fig.~S7B)\modif{, t}he  large\modif{r} space and time scales variations are encompassed by   \modif{the wave amplitude}  $\vert \widetilde{V}\vert $, with $ \widetilde{V} = \mathrm{Re}\left(\vert \widetilde{V}\vert \exp i \phi_V\right)$.
 \modif{The wave amplitude $\vert \widetilde{V}\vert $} tends to increase with \modif{the cell velocity}  $\bar{V}$  (Fig.~S7A in the Supporting Material), and accordingly decrease with \modif{the cell density} $\bar{\rho}$.

The local phase $\phi_V$ in turn determines \modif{by differentiation} the local  \modif{angular frequency} $\omega$ and wavenumber $k$. The wave velocity $c= \omega/k$ is negative here because wave and cell velocities are in opposite directions while we have chosen the convention that $V>0$. Hence $\omega$ and $k$ are of opposite signs, and with our convention $\omega >0$ while $k<0$. 
The local time period is $T= 2\pi/\omega$ and the local wavelength is $\lambda = 2\pi/\vert k \vert$. 

\modif{We observe (Figs.~\ref{fig:kymo_wt}A, S6A) that $k$ varies in  space; conversely, at a given time, $\omega$ does not vary significantly with space. Accordingly, $\omega$  does not depend explicitly on local density, which varies significantly with space.}
Interestingly, comparing experiments \modif{with different  density profiles  shows that, in a 180~min $\times$ 528~$\mu$m bin near the moving front, the wave frequency decreases  with the cell density}
(Fig.~S3B).
\modif{As already observed qualitatively (Fig. \ref{fig:kymo_drugs}A),} $c$  is of order of minus ten times the cell velocity \modif{$\bar{V}$} (compare Figs.~S3A,C) and decreases with mean effective cell radius $R_\mathrm{mean}$ (Fig.~S3C), i.e. $\vert c \vert$ decreases with $\bar{\rho}$.

Again using wavelet analysis we define and measure   $\vert \widetilde{\rho} \vert$ and $\phi_\rho$:  $\widetilde{\rho} = \mathrm{Re}\left(\vert \widetilde{\rho}\vert \exp i \phi_\rho\right)$. The kymograph of $\phi_\rho$ shows that the wavelets  detect the signal which physically corresponds to the wave velocity (Fig.~S7B). 

\modif{As already observed qualitatively (Fig. \ref{fig:kymo_drugs}A), i}nhibiting lamellipodia formation significantly decreases $\omega$ (Fig.~S3B).

\section{Phenomenological description}
\label{sec:model}

Continuum mechanics  \cite{Tlili2015} has been successfully used to model collective migration and wound closure of a cell monolayer on a substrate \cite{Lee2011,Arciero2011}.
Several models have been proposed to explain the instability which gives rise to waves by invoking one of various active cell ingredients, within the constraints raised by symmetry considerations \cite{Serra-Picamal2012,Banerjee2015,Notbohm2016,Recho2016,BMercader2017,Yabunaka2017PRE, Yabunaka2017softmatt}. 

Building on these models, we propose a simple \modif{phenomenological description (which means we model the phenomena without explicitly modeling their microscopic or biochemical causes). An} advancing front cell pulls on its follower which becomes polarized after some time delay, enabling signal propagation from the front back\modif{wards} into the bulk; when the follower eventually increases its velocity, the front cell is free to increase its velocity too, generating time oscillations. 

Our goal is to  perform testable predictions, compare them with experiments, and extract the values of relevant physical parameters.  \modif{Continuum mechanics helps here to understand the physical effect of each parameter and draw a phase diagram. Numerical simulations, which could turn useful for instance to vary boundary conditions or to link cell-level ingredients with collective migration, are beyond the scope of the present work.}

We first recall cell number and momentum conservation laws  within continuum mechanics, here expressed in one dimension (Section \ref{sec:conservation}), and couple them with the monolayer mechanical properties (Section \ref{sec:mono_mech}).
 We then include an active force linked to cell polarization (Section \ref{sec:active}), and a phenomenological coupling between strain and polarity (Section \ref{sec:coupled_eq}), to explain the existence of waves (Section \ref{sec:onset}) and predict their characteristics (Section \ref{sec:charact}).

\subsection{Conservation laws}
\label{sec:conservation}

In absence of cell division, the cell number conservation is expressed as
\begin{equation}
\frac{\partial \rho}{\partial t} + \frac{\partial (\rho V)}{\partial x} =0 
\label{conservation}
\end{equation}

\modif{For simplicity we develop a local model, neglecting the large scale gradients. W}e introduce the \modif{angular frequency} $\omega$, \modif{the} wavenumber $k<0$, and \modif{the} wave velocity $c= \omega/k <0$\modif{, and we treat them as numbers rather than as fields.}
As explained above, we separate the velocity into its average $\bar{V}$ and its variations $\widetilde{V}$, the density  into its average $\bar{\rho}$ and its variations $\widetilde{\rho}$\modif{, and again we treat $\bar{V}$ and $\bar{\rho}$ as numbers rather than as fields. Within this approximation, wavelet analysis and Fourier analysis become indistinguishable.}

We linearize Eq.~\eqref{conservation} for small wave amplitude, i.e. neglecting the $ \widetilde{\rho}\widetilde{V}$ term.
It writes $\omega \widetilde{\rho} - k ( \bar{V} \widetilde{\rho} +  \bar{\rho} \widetilde{V})=0$, or equivalently after division by $k$:
\begin{equation} 
\widetilde{\rho} \left(c- \bar{V}\right)  = \bar{\rho} \widetilde{V}
\label{rho-v} 
\end{equation}
Since the order of magnitude of $c$ is $-10\; \modif{ \bar{V}}$ (Figs.~S3A,C), Eq.~\eqref{rho-v} 
predicts that $\widetilde{\rho}$ is in phase opposition with $\widetilde{V}$, and that $\widetilde{\rho}/\bar{\rho}$ is  \modif{of order of} $- 0.1 \; \widetilde{V}/\bar{V}$.
This  explains why density oscillations  are barely visible (Figs. \ref{fig:kymo_wt}C, S5C, S6B). By measuring the  velocity and density wave characteristics at several points, we observe a local variability, which we exploit to check over a wide range that Eq.~\eqref{rho-v}  is compatible with the observed oscillation amplitudes (Fig.~S8A in the Supporting Material): Eq.~\eqref{rho-v} is checked with 10\% precision.  Phases of $\widetilde{\rho}$ and $\widetilde{V}$ should differ by $\pi$,  according to Eq.~\eqref{rho-v}, which is checked up to better than 0.03 rad, or 1\% precision (Fig.~S6D, Fig.~S7B, Fig.~S8B).

We now turn to the momentum conservation law, namely the force balance. 
The force equilibrium of the monolayer (integrated along the normal to the substrate) relates the external force per unit area $F$, exerted by the substrate on the cell monolayer, and the internal forces, namely the divergence of stress, as 
\begin{equation}  
\frac{\partial \left( h \sigma \right)}{\partial x} + F=0
\label{eq:equilibrium} 
\end{equation} 
\modif{Here} $\sigma$ is the 1D stress (equivalent to the  3D stress component along $xx$) averaged over the monolayer thickness $h$.
\modif{For simplicity, displacement and stress fields are assumed to be functions of $x$ and $t$ only and we consider only one component of each field in state equations, the component along the $x$ direction. In a real 2D description of stress,}  this would have to be replaced by the deviator of the stress tensor; alternative possibilities exist, such as the mean of the two principal stresses within the cell monolayer, i.e. half the trace of the stress tensor \cite{Notbohm2016}.

\subsection{Monolayer mechanical properties}
\label{sec:mono_mech}

In principle, the dissipation could be of both intra- or inter-cellular origin, and contribute to stress both in series or in parallel with elasticity \cite{Tlili2015}. 
These  different monolayer rheological properties are compatible with the appearance of waves \cite{Yabunaka2017softmatt}, and it is beyond the scope of the present paper to enter in such detailed description.  
To fix the ideas, the monolayer is often described as a viscoelastic liquid \cite{Lee2011,Serra-Picamal2012}, with a dissipative contribution in series with the elasticity (Maxwell model) and an elastic strain which relaxes over a viscoelastic time $\tau$:
\begin{eqnarray}
\frac{{\mathrm d}e}{{\mathrm d}t} +\frac{e}{\tau} & =& \frac{ \partial V }{\partial x}
\label{eq:relaxation}  \\
\sigma&=& Ge
\label{eq:elasticity}
\end{eqnarray}
\modif{Here, in such a 1D Maxwell model, the velocity gradient $\partial V / \partial x$ (plotted in Fig. \ref{fig:kymo_wt}B) is equivalent to the total strain rate, which in turn is the sum of the elastic and viscous strain rates. They are in series, and the viscous strain rate is $e/\tau$, where $\tau$ is the viscoelastic time and $e$ the elastic strain. It would be beyond the scope of the present paper to relate sub-cellular ingredients with  this elastic strain $e$, which we consider here as an effective, coarse-grained variable \cite{Vincent2015}. T}he elastic strain rate is ${\mathrm d}e/{\mathrm d}t = \partial e/\partial t + V\partial e/\partial x$; $G$ is the elastic modulus typically in the range $10^2-10^3$~Pa, obtained for single cell \cite{Moeendarbary2013}, by micro-indentation \cite{Harris2011,Pietuch2013} or on a monolayer \cite{Vedula_NAtMat2014} 
(note that stretching a suspended monolayer, including cell-cell junctions, yields a much larger value $\sim$~2~10$^4$~Pa  \cite{Harris2012}); the value of $\tau$ is discussed below.
From these orders of magnitude, we predict that  \modif{detecting} waves of traction force \modif{should be technically challenging}.

\subsection{Active mechanical ingredients}
\label{sec:active}

In the literature, the  force per unit area exerted by the substrate on the monolayer is often expressed as the sum of active and friction contributions, \modif{for instance} $F= f_a p - \zeta V$ \cite{Cochet2013}. 
Here $f_a$ is the characteristic value of the active force  a cell can exert; it is of order of $300$~Pa \cite{Trepat2009}, and
decreases with $\rho$ \cite{Puliafito2012,Angelini2010,Doxzen2013}. 
The dimensionless real number $p$ is \modif{a mathematical term reflecting, within the current simplified one-dimensional description, the actual two-dimensional} cell polarization.
It is convenient to introduce $V_a=f_a/\zeta$, which corresponds to a characteristic \modif{scale of} active migration velocity, and write
\begin{equation} 
\frac{F}{\zeta}= V_a p - V
\label{eq:active_decomposition} 
\end{equation}
Note that, alternatively, it would have been possible to consider $p$ as a Boolean variable, being either $+1$ or $-1$, while $V_a$ and $f_a$ \modif{would be} continuous variables. This alternative could be important when discussing for instance how  the polarity $p$ is related with biochemistry, and whether it could change sign by passing continuously through 0; but such debate is beyond the scope of the present paper. 

 \modif{To fix the ideas, w}e use  the values of the friction coefficient
 $\zeta \sim 10^9$~N~m$^{-3}$~s \cite{Cochet2013,Notbohm2016}. 
 For a wavenumber $k\sim 10^4$~rad~m$^{-1}$, and with upper estimates of the 3D cell viscosity $\eta$ of order of $10^2$~Pa~s \cite{Guevorkian2010,Stirbat-Mgharbel-2013}, 
we obtain that the modulus of the internal viscous force $\eta k \vert \widetilde{V}\vert $  is at least thousand times smaller than that of the typical external friction force, 
$ \zeta \bar{V}$.
We thus neglect the viscosity contribution in parallel with the elasticity \cite{Notbohm2016}.

Combining Eqs.~(\ref{eq:equilibrium},\ref{eq:elasticity},\ref{eq:active_decomposition})  to eliminate $F$ and $\sigma$ yields $\partial_x \left( Ghe \right) + V_a p = V$. 
Differentiating it and eliminating $\partial_x V$ with Eq.~\eqref{eq:relaxation}  yields a second order differential equation in $e$:
\begin{equation}
\frac{{\mathrm d}e}{{\mathrm d}t}  = \frac{\partial e}{\partial t} + V\frac{\partial e}{\partial x} =  \frac{\partial^{2} \left(De \right)}{\partial x^2} +  \frac{\partial \left(V_a p\right)}{\partial x} -\frac{e}{\tau}
\label{eq:e:diffusion_tau}
\end{equation}
\modif{A typical range of variation of $h$ is from 13 $\mu$m far from the front to 8 $\mu$m near the front (see Supp. Fig.~S7 of Ref.~\cite{Serra-Picamal2012}, where the cell volume is approximately conserved).}
Since $h$ varies slowly with space, Eq.~\eqref{eq:e:diffusion_tau} is locally a diffusion-like equation \cite{Bonnet2012},    \modif{where the effective strain} diffusion coefficient \modif{is} $D=Gh/\zeta$; $D$ increases with monolayer stiffness and decreases with friction. In principle, the steady flow can become unstable, and waves appear,
 if the  heterogeneity of the active term $V_ap$ overcomes the stabilising diffusion term \modif{$D$}.
A heterogeneity in migration force might create a heterogeneity in velocity,  affecting in turn the stress, which would feed back on the force. The question is how this feedback could become positive \modif{and strong enough to make the flow unstable}.

\subsection{Strain-polarity coupling equation}
\label{sec:coupled_eq}

Polarity can couple to cell strain through a mechano-sensitive protein or protein complex (such as Merlin  \cite{Das2015}).
Here we assume that  the monolayer is already polarized, symmetry is broken due to the migrating front (hence symmetry constraints \cite{Yabunaka2017PRE} are not enforced in the following equations). Cells already have a polarity $p$, which is enhanced by cell stretching and decreased by cell compression.
We neglect: non-linearities; viscosity; and the large scale variation of 
$\rho$  and $\bar{V}$ over the whole strip length scale. 
These simplifying hypotheses can easily be relaxed if required, for instance if future experiments add new details. We have checked {\it a posteriori} that a complete treatment which includes the space variations of $\rho$  and $\bar{V}$ modifies the present predictions of $\vert c \vert$ and $\omega$ by less than 10~\%.

We study the stability of a homogeneous steady  state where all cells migrate in the same direction,  $\bar{V}=V_a>0$, and are positively polarized, $\bar{p}=1$. The density is  $\bar{\rho}$, the traction force $ \bar{F}$.
We study for instance the case $\bar{e}=0$, which is relevant in the region close to the front where the waves are most visible, and which is the value towards which $e$ relaxes when $\tau$ is finite.
Note that the \modif{mirror-reflected}  steady state, \modif{where} $\bar{V}$ and $\bar{p}$ \modif{would be} negative, is irrelevant here (unlike in symmetric migration experiments \cite{Serra-Picamal2012}).
The third steady state, $\bar{V}=0$, exists initially, but ceases to be stable when the confinement is removed, and is also irrelevant 
here.

The polarity follows the strain with a delay $\tau_p$ \modif{reflecting, within the current simplified one-dimensional description, the actual two-dimensional amplitude and orientation relaxation}  time  \cite{Lee2011}:
\begin{equation}
\frac{\partial p}{ \partial t } = \frac{1+me-p}{\tau_p}
\label{eq:reorient}
\end{equation}
Here $m$ is a non-dimensional factor coupling polarity and cell strain, and when $me$ is of order 1  the polarity value changes by one unit. 
If we integrate  Eq.~\eqref{eq:relaxation} in time, using the observed velocity wave \modif{characteristics, we find} that the total strain has an amplitude of order 0.1. It means that the elastic strain has an amplitude of at most 0.1, and probably around half of it if ${\mathrm d}e/{\mathrm d}t$ and $e/\tau$ are comparable (which is the case since $\tau$ is comparable with the wave period, see below). Hence a strain  of at most 0.1 (and possibly 0.05) suffices to change the polarity from value 0 to value 1, and $m$ is of order of  10, possibly 20, or 25 at most.  

\subsection{Onset of wave appearance}
\label{sec:onset}

To perform a linear stability analysis around the steady state $\modif{\bar{e}}=0$, $\modif{\bar{p}}=1$, the small variables are $\delta e=e$ and $\delta p= p-1$. \modif{T}erms due to variations of $h$ are \modif{of} second order and thus negligible. 
Eqs.~(\ref{eq:e:diffusion_tau},\ref{eq:reorient}) become, after linearization:
\begin{eqnarray}
\frac{{\mathrm d} \delta e}{{\mathrm d} t}  &=&  D \frac{\partial^{2} \delta e}{\partial x^2} +  V_a  \frac{\partial \delta p}{\partial x} -\frac{\delta e}{\tau}
\nonumber \\
\tau_p \frac{{\mathrm d} \delta p}{{\mathrm d} t } &=&  m\delta e-\delta p 
\label{eq:linear_coupled}
\end{eqnarray}
 
We look for small perturbations (of the steady, homogeneous state) 
proportional to $\exp\left(st-ikx\right)$, where the wavenumber $k$ is a real number, and the wave growth rate $s$ is a complex number with a real part which is strictly positive when the steady state is unstable, $\mathrm{Re}(s)>0$, and  a non zero imaginary part, $\mathrm{Im}(s) \neq 0$.
The Jacobian matrix of the  equation system, Eqs.~\eqref{eq:linear_coupled}, is:
\begin{eqnarray}
{\begin{pmatrix}
  s-ik\bar{V}+Dk^{2}  + \tau^{-1} & ikV_a \\
 - m& \tau_p \left(s-ik\bar{V}\right)+1 \\
  \end{pmatrix}}
\label{eq:jacobian_2D}
\end{eqnarray}
  
Solving in $s$ simply requires to write that the determinant of ${\cal J}$ is zero:
\begin{equation}
\tau_p S^2 + \left( 1 + \tau_p Dk^{2} + \frac{\tau_p}{\tau} \right) S  + Dk^{2} +  \frac{1}{\tau} + imkV_a = 0
\label{eq:determinant}
\end{equation}
where $S=s-ik\bar{V}$.
There are two roots, which depend on $k$ and on parameter values. 
We numerically  solve Eq.~\eqref{eq:determinant} and look for a root with $\mathrm{Re}(s)>0$ and   $\mathrm{Im}(s) \neq 0$.
Depending on the parameter values (Fig.~\ref{fig:diagphase}), there exists a range of $k$ with one such root, and the steady, homogeneous solution is unstable. A propagating wave appears; the mode $k$ which develops more quickly is the one for which $\mathrm{Re}(s)$ is maximum, i.e. ${\mathrm d} \mathrm{Re}(s)/{\mathrm d}k=0$  (until the amplitude  increases enough to reach the non-linear regime). Its imaginary part $\mathrm{Im}(s)$ is the \modif{angular frequency} $\omega$ of this mode. 

To fix the ideas and provide example of calculations, we take  from experiments that $\bar{V}$ is of order of \modif{1} $\mu$m.min$^{-1}$. The delay time $\tau_p$ of polarity with respect to stretching, due to the reaction time of \modif{the} Rac \modif{pathway}, could be of order of 25~min \cite{Das2015}. 
The value of $m$\modif{, ranging from 10 to 25,} also affects the predictions; higher $m$ values tend to make the absolute value of $c$ larger ($c$ is more negative).
We  defer the discussion of $V_a$ and $D$ to \modif{Section \ref{sec:charact} and} now discuss the value of $\tau$.

In the viscoelastic liquid description we consider here, at time scales smaller than $\tau$ the monolayer can sustain a shear stress and behaves as mostly solid, while at longer time scales the strain relaxes and the monolayer behaves as mostly liquid. So it is important to determine whether $\tau$ is larger or smaller than the time scale of the waves, \modif{which is} of order of one hour. This is subject to debate, since depending on the cell line the elastic modulus and the viscoelastic time of tissues can vary over orders of magnitude \cite{Vincent2015}. Even when restricting to MDCK monolayers, published values for the viscoelastic time $\tau$  range from 15~min \cite{Lee2011} to 3 - 5 hours \cite{Vincent2015}. Several articles  \cite{Trepat2009, Arciero2011,Serra-Picamal2012}  choose to treat the monolayer as elastic, given that the elastic modulus can be an effective modulus arising from the cell activity \cite{Hawkins2014,Vincent2015}. 

We have checked numerically  that small values of $\tau$, for instance 30 min or less, stabilize the steady state, while large values of $\tau$, 1 hour or more,  allow for the wave appearance. The wave characteristics we determine barely change when $\tau$ spans the range 1 to 5 hours.

\subsection{Wave characteristics}
\label{sec:charact}

\modif{The active cell velocity} $V_a$ makes waves appear while \modif{the strain diffusion coefficient}  $D$ is damping them. Experimental measurements of wave characteristics can help estimate orders of magnitude of $D$ and $V_a$. However, determining their precise values is difficult and strongly dependent on the model (which is here only phenomenological). 
We thus let values of $D$ and $V_a$ vary within a reasonable range and solve systematically Eq.~\eqref{eq:determinant}, to determine a phase diagram in the $(D,V_a)$ plane.

\begin{figure}[h!]
\noindent (A) \hfill   (B) \hfill  ~  \\
\showfigures{\includegraphics[width=0.476\columnwidth]{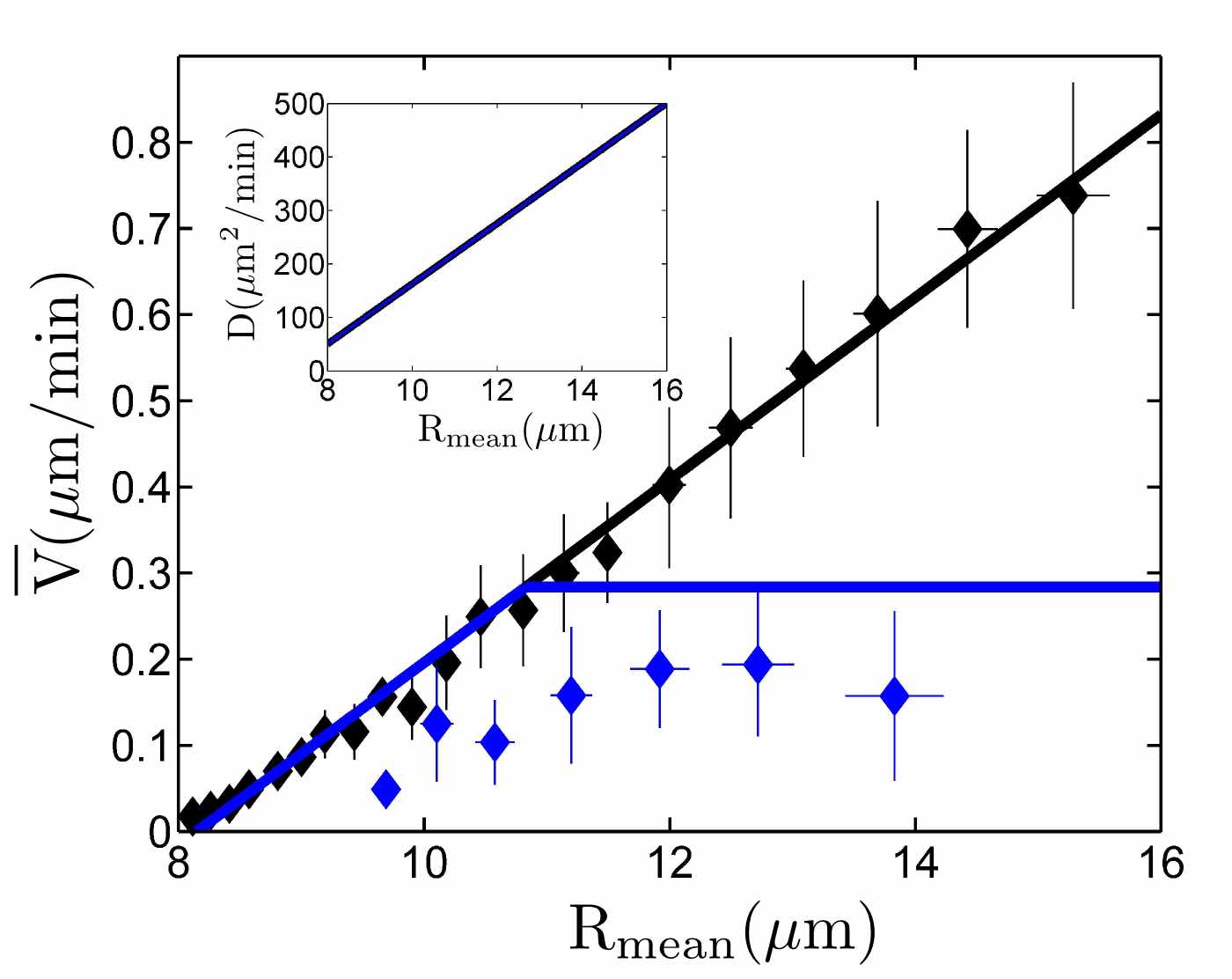}} 
\showfigures{\includegraphics[width=0.499\columnwidth]{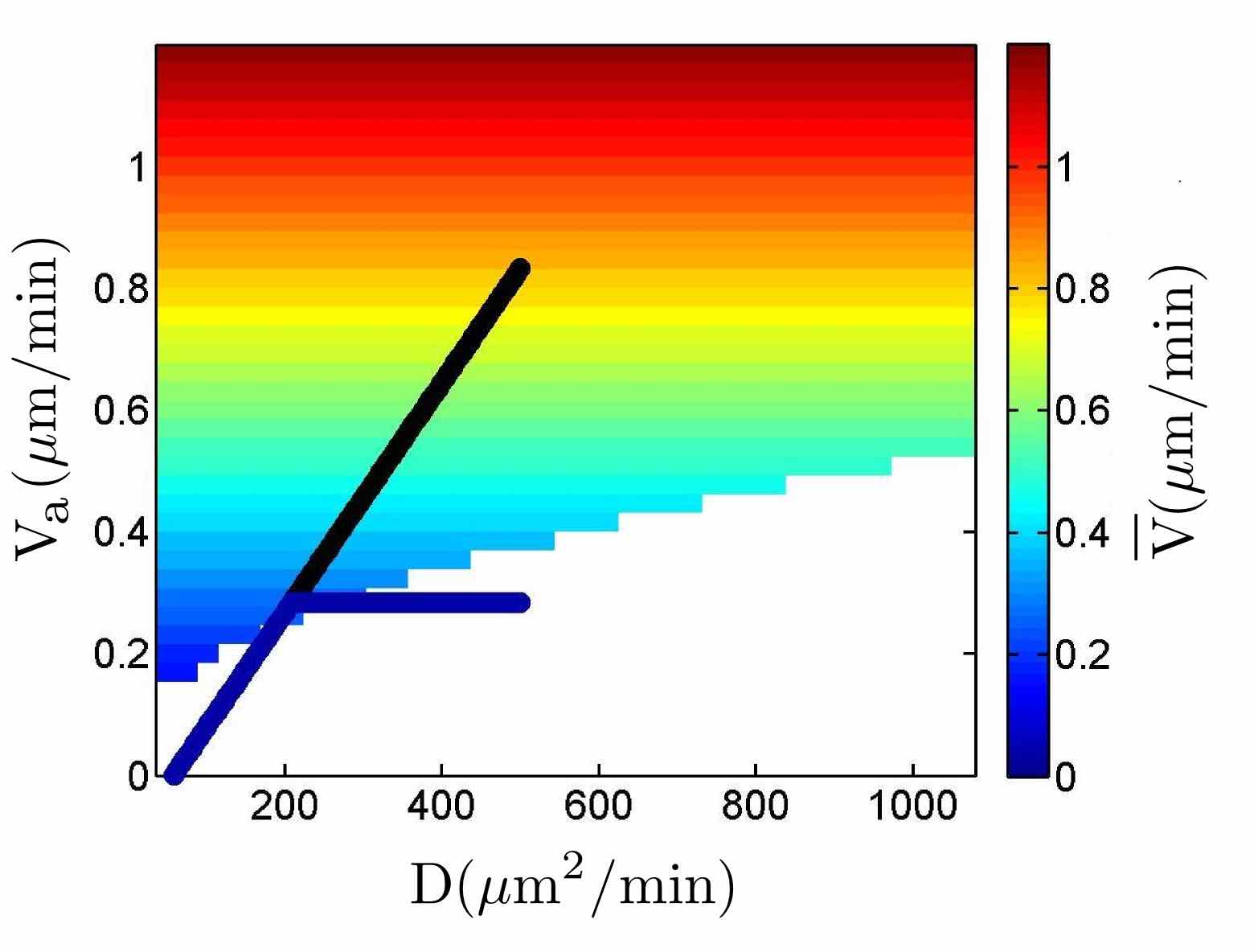}} 
\\
\noindent (C) \hfill   (D)   \hfill  ~  \\
\showfigures{\includegraphics[width=0.476\columnwidth]{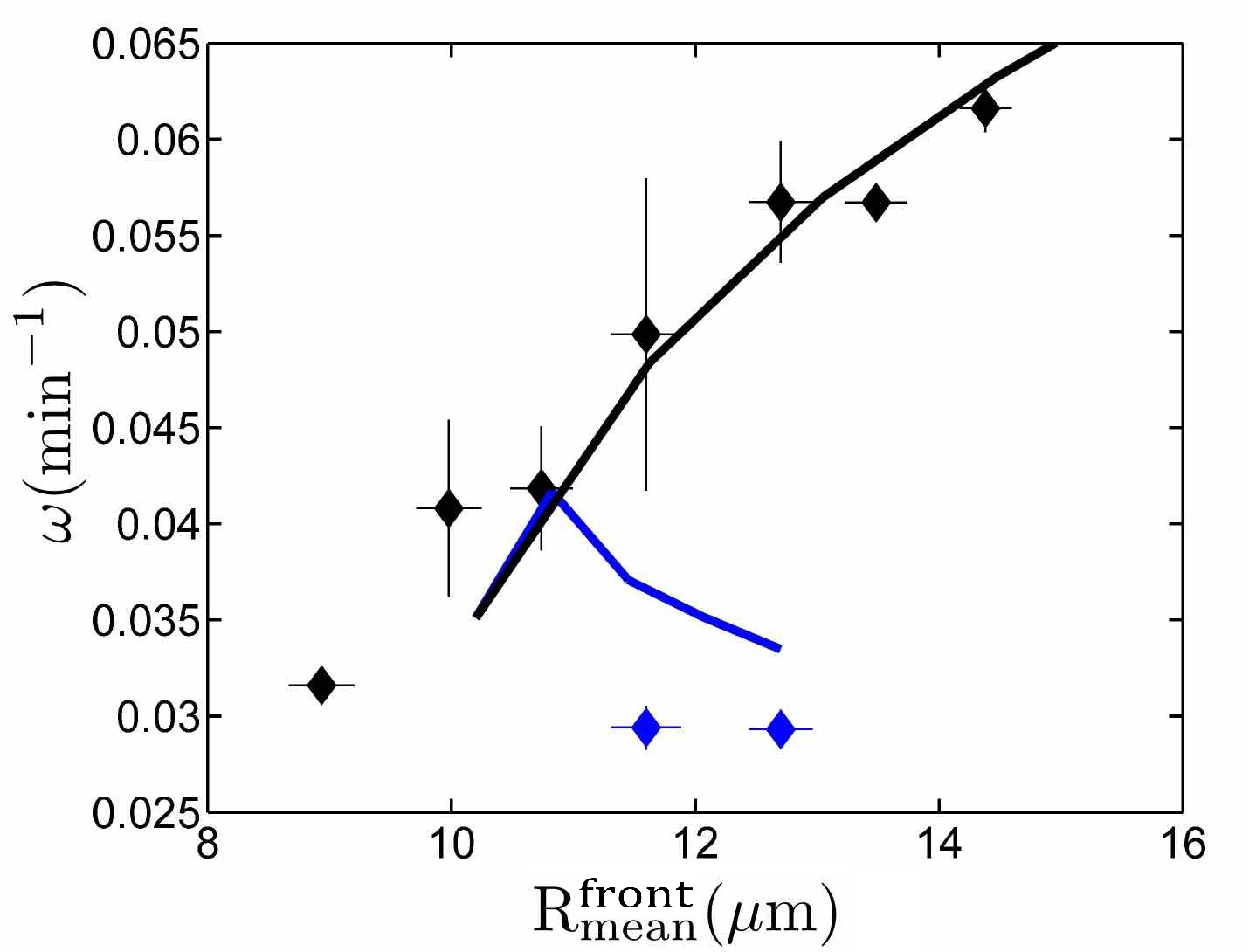}} 
\showfigures{\includegraphics[width=0.499\columnwidth]{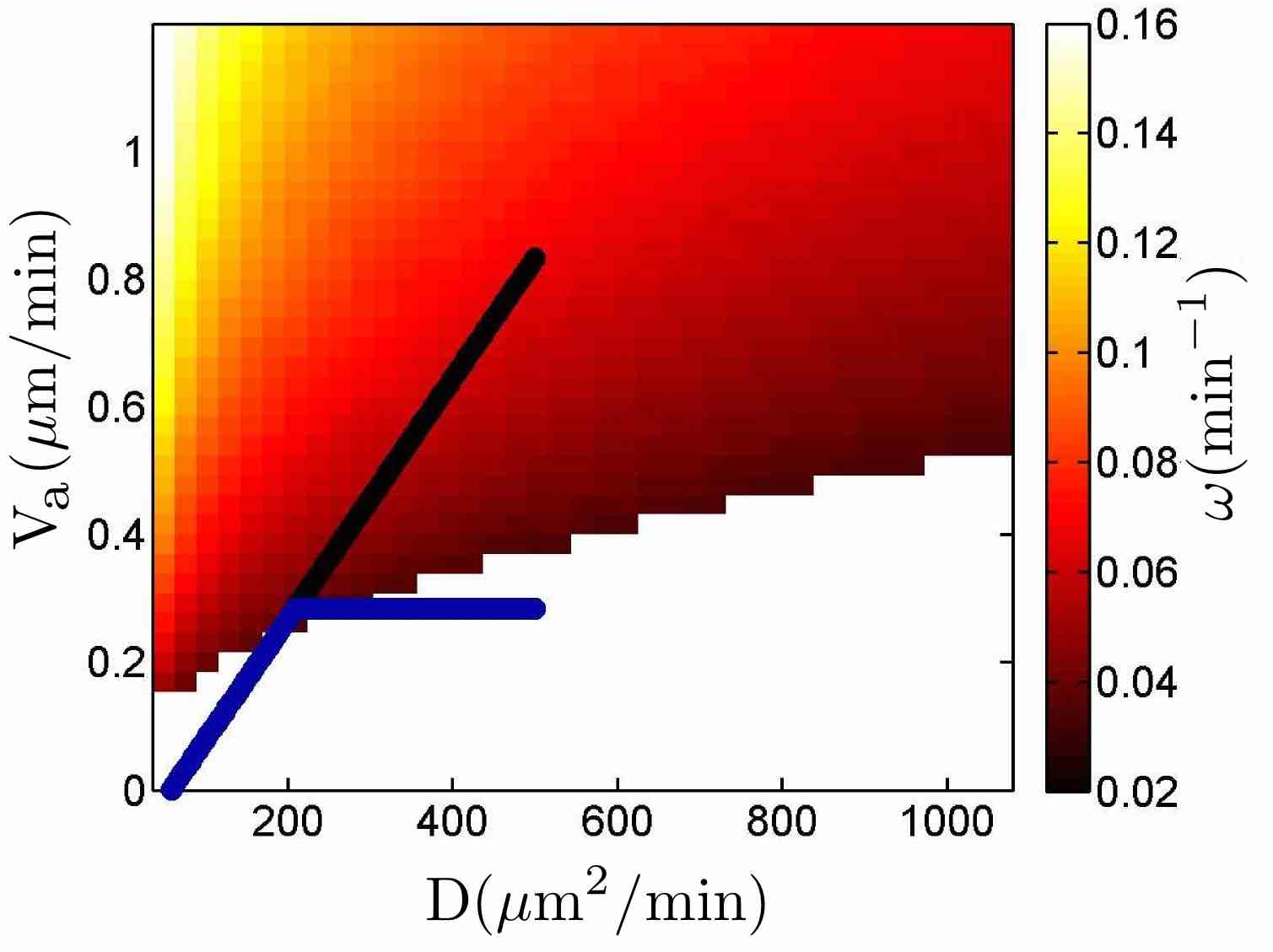}} 
\\
\noindent (E) \hfill   (F) \hfill    ~  \\
\showfigures{\includegraphics[width=0.476\columnwidth]{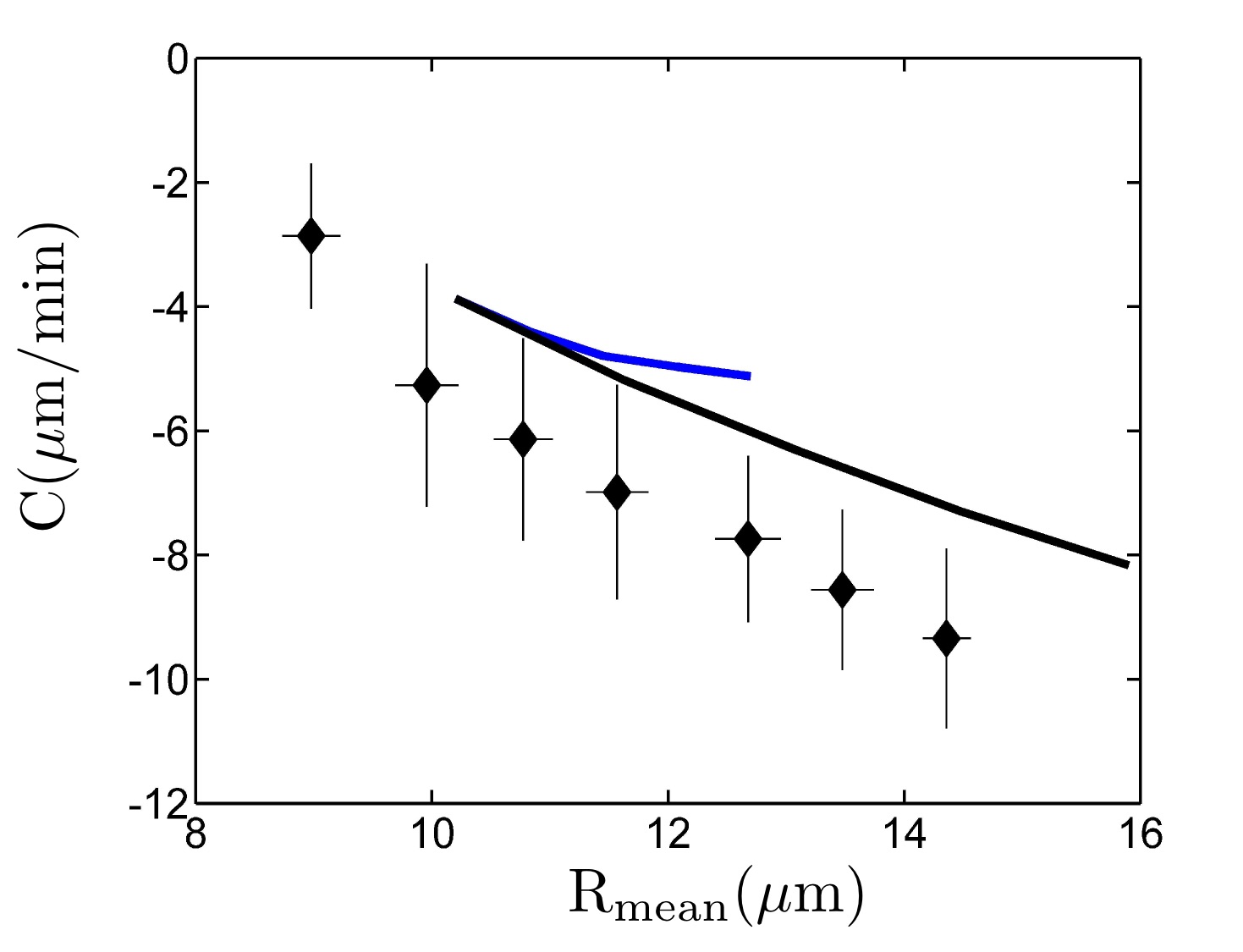}} 
\showfigures{\includegraphics[width=0.499\columnwidth]{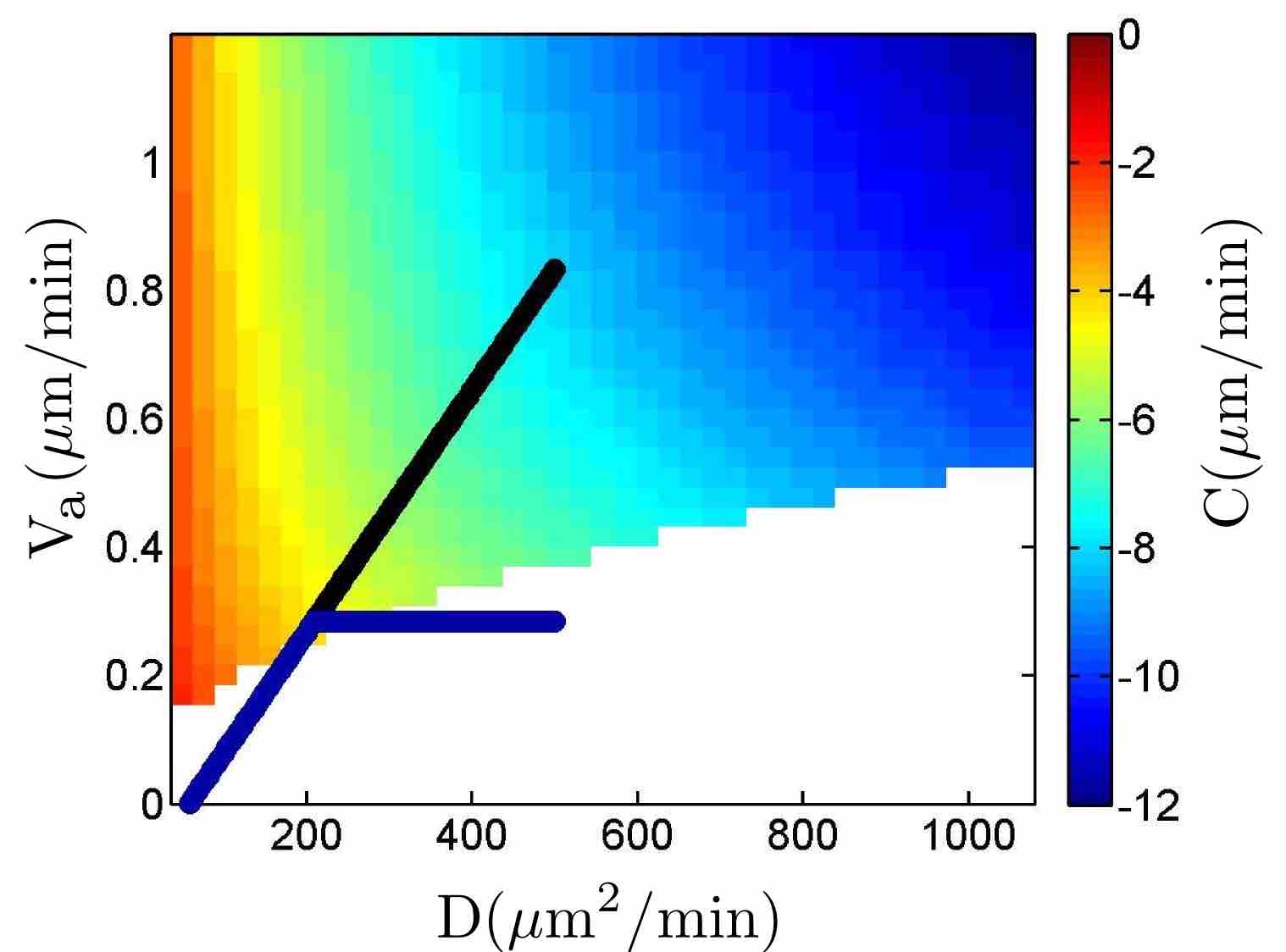}} 
\caption{
Predictions of phenomenological model. \modif{Realistic p}arameter values (\modif{strain-polarity coupling term} $m = 25$, \modif{polarisation delay time} $\tau_p = 15$~min, \modif{viscoelastic time} $\tau = 180$~min)  are manually chosen to obtain a good agreement with the data, see text. 
{\it (A)} \modif{Large-scale average of cell velocity} $\bar{V}$ vs mean effective cell radius $R_{\mathrm{mean}}$. 
Points are experimental data \modif{from Figs.~\ref{fig:profiles}C,D}. Black:   experiments with mitomycin; \modif{black line: linear relation}. Blue:  experiments with lamellipodia inhibition using CK666  \modif{drug; blue line:} we draw a linear increase followed by a plateau (see text for details). Inset: $D$ vs $R_{\mathrm{mean}}$, estimated relation, not affected by CK666.
{\it (B)} Diagram of cell velocity $\bar{V}$ vs \modif{strain diffusion coefficient $D$ and active velocity $V_A$}. 
It is plotted in the region where the wave amplitude growth rate is positive (existence of waves). The regions where the wave amplitude growth rate is negative (the steady migration is stable) are left blank. 
Lines correspond to those in (A); $R_{\mathrm{mean}}$ is increasing from bottom left to top right.
{\it (C, D)} Same for wave \modif{angular frequency $\omega$, which depends on the mean effective cell radius; they are  measured  in a 180~min $\times$ 528~$\mu$m \modif{bin} near the moving front (Fig.~S3B)}.
{\it (E, F)} Same for wave velocity $c$; note that  its values are negative, and that with CK666 \modif{drug} the  values of $c$ are too noisy for quantitative measurements, but are similar. 
\label{fig:diagphase}
}
\end{figure}

In Fig.~\ref{fig:diagphase} we plot the model predictions for \modif{manually chosen, realistic} parameter values $\tau_p = 15$~min, $\tau = 180$~min, and $m = 25$\modif{. T}he experimental data  indicate that both $V_a$ and $\vert c \vert $ increase linearly with $R_{\mathrm{mean}}$ (Fig.~\ref{fig:profiles}C; Fig.~\ref{fig:diagphase}E, Fig.~S3C). 
We use the measured relation
 $\bar{V} = 0.106 \; R_\mathrm{mean} - 0.864$ 
(Fig.~\ref{fig:profiles}D; Fig.~\ref{fig:diagphase}A).
To reproduce the experimentally observed  linear variation of $\vert c \vert$ with  $R_{\mathrm{mean}}$ (Fig.~\ref{fig:diagphase}E), we find that  $D$ too has to increase with $R_{\mathrm{mean}}$; in the following we choose a linear relation between    $D$  and $R_{\mathrm{mean}}$ (inset of Fig.~\ref{fig:diagphase}A).
This determines the black lines on Fig.~\ref{fig:diagphase}B-F as possible variations of $\bar{V}$, $c$ and $\omega$, and corresponding trajectories in the $(D,V_a)$ plane,  when $R_{\mathrm{mean}}$ varies.
We obtain an agreement with measurements of \modif{$\omega$ versus $R_{\mathrm{mean}}$ at the front}, which we reproduce quantitatively \modif{(Fig.~\ref{fig:diagphase}C), and of $c$ versus $R_{\mathrm{mean}}$}, which we capture qualitatively \modif{(Fig.~\ref{fig:diagphase}E)}. 
Note that $c$ is negative, indicating backward waves as observed in experiments. 

The instability threshold, visible as the limit between colored and blank regions  (Fig.~\ref{fig:diagphase}B,D,F), indicates when $V_a$ is strong enough to overcome $D$. We obtain a consistent picture with typically $D$ of order of $10^2$~$\mu$m$^{2}$/min (i.e. $G$ of a few $10^2$~Pa), $V_a$ of order of $10^{-1}$~$\mu$m/min ($f_a$ \modif{of order} of one or a few $10$~Pa),  wavelength of several $10^2$ or $10^3$~$\mu$m, time period of a few $10^2$~min, $\vert c \vert$ of order of $10^1$~$\mu$m/min.

We observe that CK666 \modif{drug treatment} results in a saturation in $\bar{V}$, probably linked with a limit in lamellipodia size \cite{Vitriol2015} (Fig.~\ref{fig:profiles}C). 
To position the corresponding predictions (blue lines in Fig.~\ref{fig:diagphase}) without introducing free parameters, we proceed as follows. 
Since within the model $V_a=\bar{V}$, we use at small $R_{\mathrm{mean}}$ values  the same linear increase as without \modif{CK666} (Figs.~\ref{fig:profiles}D; Fig.~\ref{fig:diagphase}A), and at larger $R_{\mathrm{mean}}$ values  the saturation corresponds to a plateau.
To determine the position of the cross-over between these regimes, i.e. the onset of saturation, we observe that  \modif{in experiments with CK666  drug,} the monolayer is very close to the limit of wave appearance: \modif{it can  lead either to decrease in wave amplitude and frequency, or to a suppression of the wave}. We thus position the cross-over at the intersection between the straight line and the instability threshold; this qualitatively captures all features of Fig.~\ref{fig:diagphase} without \modif{additional} free parameter. 

Experimentally, in the case where small waves are observed in presence of CK666  \modif{drug}, then $R_{\mathrm{mean}}$ is near 12~$\mu$m, $\omega$ is near 0.03~min$^{-1}$. In this case the wave velocity $c$ measurement is too noisy to be quantitative, but $c$ seems unaffected as no slope rupture is visible on the \modif{wave pattern} (Fig. \ref{fig:kymo_drugs}A). These features are qualitatively captured by our model.

\section{Discussion}
\label{sec:discussion}

We first discuss the average velocity profile and show that \modif{it depends on density rather than on distance to moving front}; our quantification evidences the effect of proliferation and the key role played by polarity in cell migration (Section \ref{sec:cellmig}), as well as in velocity waves  
(Section \ref{sec:waveres}). Our experiments and our phenomenological model agree reasonably, thus improving our understanding (Section \ref{sec:exppheno}), and contributing to explain recent observations of  backward waves in colliding monolayers \cite{RodriguezFranco2017}.
 More broadly, our experimental data help \modif{to} discriminate between existing theories (Section \ref{sec:comparemod}). 
	
\subsection{Cell migration results}
\label{sec:cellmig}

Experiments had shown that $V$ decreased with the distance \modif{$d$}
to the moving front \cite{Serra-Picamal2012} while $\rho$ and $\sigma_{xx}$ increased with \modif{$d$} and were proportional to each other \cite{Trepat2009}. We can imagine two possible interpretations: either that both $\rho$ and $\sigma_{xx}$ happen to vary similarly with \modif{$d$}, with a reinforcement of cell-cell junctions from the front to the back; or  that  $\sigma_{xx}$ is actually determined by $\rho$.

Here, we observe a large enough range of cell densities \modif{$\rho$ and velocities $V$, and of distances  $d$ to the front}, with a good enough signal-to-noise ratio, to discriminate between \modif{$V$ depending on $d$ versus on $\rho$}. 
We find that in the monolayer bulk,   $V$ depends only on $\rho$, namely that it increases  linearly with $R_{\mathrm{mean}}$, irrespectively of \modif{$d$} or of the past history of the cell monolayer  (advection, divisions, extrusions) which causes the observed density value.
This is compatible with observations of Refs. \cite{Serra-Picamal2012,Trepat2009}; it suggests that  the traction force is cell-autonomous and is linear in $R_{\mathrm{mean}}$;
and it favors \modif{the} interpretation that  $\sigma_{xx}$ is determined by $\rho$.

\modif{If the monolayer spreading was only determined by stretching under a stress gradient, then it would be similar to a passive material wetting a solid substrate; the velocity and density profiles would depend on the distance to the front. Our experiments rule out this description: indeed, the velocity profile rather depends directly on the density, and the cell-autonomous active term of Eq.~\eqref{eq:e:diffusion_tau},  ${\partial \left(V_a p\right)/\partial x}$, plays an important role. Conversely, the stretching under a stress gradient actually feeds back on the cell polarity and activity, through Eq.~\eqref{eq:reorient} and it\modif{s} non-autonomous signalling term $me$. We suggest that this mutual coupling between cell-autonomous active migration and non-autonomous stretching by neighbouring cells, summarized by Eq.~\eqref{eq:linear_coupled}, gives rise to collective migration.}

\modif{Oriented cell divisions have two antagonistic effects  (Fig.~\ref{fig:profiles}C; Fig.~S4). First, they contribute to increase the cell movement and the front velocity; they also increase the noise in cell velocity, and more regions have a negative velocity (Fig.~S4D). 
Hence at a given density, a proliferating monolayer has a higher velocity than a non-proliferating one. Second, however, a monolayer with a high proliferation rate has a different time evolution: its density increases with time (while a monolayer with a low proliferation rate or no proliferation at all has a density which decreases with time due to spreading). A monolayer with a high proliferation rate has a velocity which decreases with time and 
reaches within hours a density where  cells are jammed and lamellipodia are absent, the front velocity is low and mainly due to divisions.}

With migration alone, \modif{in absence of division}, coherent cell collective movement propagates over several mm inside the monolayer. The wavelength is of order of 1~mm and more broadly speaking the whole velocity field is established coherently over 4 mm.
We have used 200~$\mu$m and 1~mm wide strips, larger than the typical 200~$\mu$m correlation length for the cell velocity field \cite{Poujade2007,Vedula2012}. We do not detect any significant effect on the strip width on the results presented here. 

Our observations are compatible with a large-scale polarized activity induced for instance by activity of the protein Merlin  (a tumor suppressor), as recently shown experimentally  \cite{Das2015}. The cells at the front of the migrating monolayers are known to exert large traction forces \cite{Vedula_NAtMat2014,Reffay2014} that can induce the build-up of a large intercellular stress and in turn, a polarization of the following cell by a relocalization of Merlin from the cell-cell junctions to the cytoplasm. When the cell is at rest, Merlin is localized at the cell-cell junctions. This junctional Merlin inhibits the formation of cryptic lamellipodia. On the other hand, when cell-cell junctions experience a stretching stress, Merlin is relocated to the cytoplasm. Due to the decrease in junctional Merlin, the Rac pathway becomes activated, within a delay time of a few tens of minutes. Then, within a  much smaller delay, Rac activates  the generation of cell polarization and lamellipodia, responsible for the migrating forces  \cite{Das2015}. The iteration of such processes may lead to large scale polarization within the tissue.

To complement existing experiments with blebbistatin which focus on the role of cell contractility \cite{Serra-Picamal2012,Notbohm2016},  we inhibit lamellipodia with CK666  \modif{drug treatment}. It has a clear effect, even in the bulk of the monolayer, on $V$ which is decreased; and on the $V(R_{\mathrm{mean}})$ relation, which saturates and is no longer linear (Fig.~\ref{fig:profiles}C). This suggests that the contribution of lamellipodia to the traction force is dominant, and linear in $R_{\mathrm{mean}}$. 

\subsection{Wave results}
\label{sec:waveres}

\modif{Our observations and our model agree with experimental observations by Trepat and coworkers (Fig. 3a,b of Ref. \cite{Serra-Picamal2012}, and Ref. \cite{RodriguezFranco2017}) that waves arise at the front and propagate backwards, with wave velocity direction opposed to cell velocity direction}. 

In experiments with divisions, \modif{a monolayer with a high proliferation rate leaves too quickly, or never reaches, the low density regime where large-amplitude steady waves develop. T}he mechanical waves are slightly visible and overdamped; this is broadly compatible with the literature \cite{Serra-Picamal2012,Zaritsky2014,Deforet2014,Notbohm2016}. 

Without divisions, we obtain a good enough signal-to-noise ratio to measure the wave properties. 
Moreover, thanks to this signal-to-noise ratio and experiment reproducibility, we can even measure the variation of wave properties across space and time, and with enough precision to discriminate between dependence with position vs with density.
We observe that the wave velocity $c$ is \modif{of order of} $- 10\; \modif{\bar{V}}$ and, like $\modif{\bar{V}}$, it depends explicitly only on $\rho$: it is linear in $R_{\mathrm{mean}}$, again irrespectively of \modif{distance $d$ to the moving front} or of the past history of the cell monolayer.

For a given experiment, although $\rho$ is space-dependent, the wave \modif{angular frequency} $\omega$ is spatially homogeneous \modif{(Figs.~\ref{fig:kymo_wt}A, S6A)}. This might result from the most developed mode temporally forcing the instability over the whole monolayer. As a consequence, the wavenumber $k$ depends on space.
Now, comparing experiments at different densities, we observe that $\omega$ increases  with $R_{\mathrm{mean}}$ \modif{measured near the front}  (Fig.~\ref{fig:diagphase}C, Fig.~S3B). Inhibiting lamellipodia formation decreases $\omega$, at a given \modif{value of $R_{\mathrm{mean}}$ measured \modif{near} the front (Fig.~S3B)}.

Backwards propagating waves are reminiscent of a generic instability mechanism originally discussed  in the context of car traffic \cite{Lighthill1955}, which arises when the velocity $V$ is a decreasing function of density $\rho$.
For instance, velocity pulses have been observed for dense colloids in channel flow near jamming - unjamming 
transitions, in experiments \cite{Isa2009} and in simulations \cite{Kanehl2017}.
Similarly, self-propelled agents, which tend to accumulate where they move more slowly and/or slow down at high density (for either biochemical or steric reasons) undergo a positive feedback which can lead to motility-induced phase separation between dense and dilute fluid phases \cite{Cates2015,Solon2018}. 

\subsection{Comparison of experiments with phenomenology}
\label{sec:exppheno}

Inspired by published observations and by our own, we propose here a simple description where motility forces in the bulk of a homogeneous monolayer are oriented by a dynamic pulling on cell-cell junctions. This elasticity-polarity coupling is combined with classical rheology equations of continuum mechanics. 

\modif{Our experimental observations with drugs against proliferation or lamellipodia are compatible with the theoretical picture in which w}aves spontaneously appear close to the instability limit (Fig.~\ref{fig:diagphase}). This could explain why in \modif{preceding experiments with proliferation and jamming, waves were damped and more difficult to extract from noise beyond one period}  \cite{Serra-Picamal2012,Zaritsky2014,Deforet2014} (see also the Supp. Figs. S2, S5 of Ref.~\cite{Notbohm2016})\modif{. It also explains that wave observations are sensitive to experimental details;}
parameters can change from one experiment to another depending on cell size or substrate \modif{properties such as stiffness or} coating. 

With reasonable parameters values,  typically $G$ of a few $10^2$~Pa, $f_a$ of one or a few $10$~Pa,  we find  propagating waves with wavelength of several $10^2$ or $10^3$~$\mu$m, time period of a few $10^2$~min, of the same order as the experimental values. 
We predict a negative wave velocity $c$ indicating backward waves, with  $\vert c \vert$ of order of $10^1$~$\mu$m/min,  as observed in experiments. This \modif{backward propagation can probably be explained} because a cell migrating towards the front pulls the cell behind it and favors its migration (in our model, under traction the cell polarity increases, with a positive coupling factor $m$). 

We expect that when $\rho$ decreases \modif{(}$R_{\mathrm{mean}}$ increases\modif{)}, $V_a$ significantly increases  (Fig.~\ref{fig:diagphase}). It is compatible with our observation that, when
comparing experiments at different densities,  $\omega$ decreases with $\rho$
(Fig.~\ref{fig:diagphase}C). 
Our model presents a Hopf bifurcation sensitive to the density,  and a slowing down of the wave \modif{frequency} when approaching the bifurcation. With proliferation due to cell division, the density increases; it can  lead either to decrease in wave amplitude and \modif{frequency}, or to a suppression of the wave, as observed in experiments 
(Fig.~\ref{fig:kymo_drugs}, Fig.~\ref{fig:diagphase}C). 
The model suggests that two experiments performed at a slightly different initial density can, after lamellipodia inhibition, lead either to decrease in wave amplitude and \modif{frequency}, or to a suppression of the waves; this could explain the observed effects of  CK666  \modif{drug} (Fig. \ref{fig:kymo_drugs}, Fig.~\ref{fig:diagphase}C). 

Our model qualitatively suggests that $D$ increases with $R_{\mathrm{mean}}$ (Fig.~\ref{fig:diagphase}A, inset). Our estimation of $D$ which increases linearly, by a factor of ten, when $R_{\mathrm{mean}}$ doubles (Fig.~\ref{fig:diagphase}A, inset) is compatible with the observation that the elastic modulus can vary over orders of magnitude, and scales linearly with the  size of the constituent cells  \cite{Vincent2015}. This could be compatible with the intuition that a cell which spreads has more stress fibers and a more organized cortex, resulting in a larger cell stiffness $G$. It is also compatible with the fact that the relation $D\left( R_{\mathrm{mean}} \right)$ is much less influenced than the relation $V_a\left( R_{\mathrm{mean}} \right)$ by the lamellipodia inhibition. 
Note that alternative explanations \modif{of $D$ variations with cell size} exist, for instance if the friction coefficient $\zeta$ was decreasing with  $R_{\mathrm{mean}}$. 

In summary, our model (Fig.~\ref{fig:diagphase}) is precisely compatible with measurements of $\bar{V}$, reproduces quantitatively $\omega$ and qualitatively $c$, which sign is correctly predicted. Our predictions of how $V_a$ and $D$ vary with $R_{\mathrm{mean}}$ agree with independent experiments on the same MDCK cells \cite{Vincent2015}. In presence of CK666  \modif{drug}, our model is qualitatively  compatible with either the suppression of waves, or waves with a smaller \modif{amplitude and frequency}
and an unchanged \modif{velocity}.

\subsection{Comparison with existing models}
\label{sec:comparemod}

Marcq and co-workers have shown that since cells are active, the appearance of waves is compatible with a wide range of ingredients. In particular, the rheology is not crucial, as waves can appear in materials with various rheological behaviours \cite{Recho2016,Yabunaka2017softmatt}. They predict that there are more waves when divisions are inhibited \cite{Yabunaka2017PRE}  and less waves when lamellipodia are inhibited \cite{Yabunaka2017softmatt}. Our observations and model agree with these predictions.

Banerjee, Marchetti and coworkers \cite{Banerjee2015,Notbohm2016} predict the existence of waves depending on an effective elasticity and traction force amplitude. Our model is based on an approach similar to theirs; we make it simpler while keeping the main ingredients. They find a wave velocity comparable to a cell length divided by the time ($\sim 2$~min) required for mechanical stress information to propagate across the cell. Their wave period, around 6$\pm$2~h,  increases with decreasing traction forces. 
Their predictions are consistent with three of our observations. First, their waves propagate backwards. Second, the wave \modif{frequency increases} with increasing density (and thus decreasing traction force) at the migrating front. Third, when adding CK666  \modif{drug}, either the wave \modif{frequency decreases}  or the wave disappears. 

Blanch-Mercader and Casademunt  \cite{BMercader2017}  explain that even viscous tissues can have an effective elasticity in which waves appear. This is compatible with the idea, arising from both  experiments  \cite{Vincent2015} and modeling  \cite{Hawkins2014}, that the elastic modulus can be an effective modulus arising from the cell activity. 

Trepat and coworkers present numerical simulations  \cite{Serra-Picamal2012} where, based on their cell stretching experiments,  they introduce a dynamically changing elasticity (non-linear cytoskeletal reinforcement). Our ingredients are similar to theirs except that, inspired by recent experiments on collective cell migration \cite{Das2015}, we introduce a dynamically changing traction force modulated  by the elastic strain. 

Note that  \modif{in Fig. 4d-f of Ref. \cite{Serra-Picamal2012}, simulations} predict forward-propagating waves, \modif{arising at the center and moving toward the front,} i.e. waves velocity in the same direction as cell velocity. 
\modif{Fig. 6 of Ref. \cite{BMercader2017} too predicts forward propagating waves. In contrast, experiments of Refs. \cite{Serra-Picamal2012,RodriguezFranco2017}, our experiments and model agree that waves arise at the front and propagate backwards.}
We are not aware of any interpretation of this discrepancy.

\section{Conclusion}

In summary, by inhibiting cell proliferation in a cultured epithelial cell monolayer, we limit density increase and observe steady migration over a day or more, without reaching jamming densities.  We observe for the first time a coherent collective cell migration propagate over several millimeters; cells spread and density decreases from the monolayer bulk towards the front.   Cell velocity increases linearly with cell effective radius, and does not depend directly on the distance to the front. 

On top of this average velocity profile, we detect ten periods of backwards propagating velocity waves, with millimetric wavelength. The signal-to-noise ratio is sufficient to perform precise and reproducible measurements of local characteristics  of the wave and their spatial variation. Their velocity (around ten micrometers per minute) is ten times the cell velocity; it  increases linearly with cell radius. Their period  (around two hours) increases with the cell density at the front. The period is spatially homogeneous, which might result from the most developed mode temporally forcing the instability over the whole monolayer. As a consequence, the wavenumber depends on space.
Density waves also appear, with a tiny amplitude and a phase opposed to that of velocity waves. 

The most visible effect of cell divisions is to steadily increase cell density, which contributes to jamming and decreases the migration velocity. However, at a given density,  divisions  contribute to increase  front velocity, cell velocity  and noise in cell velocity.
 When we inhibit lamellipodia formation, cell velocity drops while waves either disappear, or have a smaller amplitude and slower period. Our results suggest that the lamellipodia contribution to cell traction force is dominant \modif{and} linear in the cell radius.
 
We propose a simple model in which motility forces in the bulk of a homogeneous monolayer are oriented by a dynamic pulling on cell-cell junctions. Our model combines conservation laws, monolayer mechanical properties, and a phenomenological coupling between strain and polarity: an advancing front cell pulls on its follower which becomes polarized after some time delay (possibly through the Merlin/Rac pathway), enabling signal propagation from the front back into the bulk; when the follower eventually increases its velocity, the front cell is free to increase its velocity too, generating time oscillations. 

We find that  waves appear spontaneously  but are very close to the instability limit, which could explain why  in the past \modif{waves were damped and difficult to extract from noise beyond one period}. 
Parameter values close to the instability limit yield qualitative and quantitative predictions compatible with observations, including: waves propagate backwards; wave velocity increases with cell radius; lamellipodia inhibition attenuates, slows down or even suppresses the waves; cells maintain their polarity far from the migrating front. \modif{An interpretation of our results is}  that both cell and wave velocities depend on lamellipodia activity. 
\modif{This suggests that increasing traction forces, and/or decreasing the friction, would increase the cell and wave velocities;  increasing the monolayer stiffness, and/or decreasing the friction, would increase the strain diffusion coefficient, and thus decrease the wave amplitude and frequency.}

Together, our experiments and model  \modif{disentangle the respective contributions of polarized active velocity and of proliferation in monolayer migration}. They highlight the importance of  \modif{coupling between non autonomous strain on one hand, and autonomous polarity and migration on the other hand}, in collective cell migration and waves. They suggest that  a cell on the substrate exerts a traction force which is cell-autonomous and linear in the cell radius, and that the ratio of cell stiffness to  cell-substrate friction increases with cell radius. \modif{Finally, they reveal the central role of density in determining cell and wave velocities.}

\vskip1pc

\ethics{NA}

\dataccess{Figures supporting this article are provided in the Supplementary material. Movies, code and measurements supporting this article are available from the Dryad Digital Repository \cite{Tlili2018Dryad}: \url{https://doi.org/10.5061/dryad.sk512}}

\aucontribute{
S.T.,
B.La.,
H.D.-A.,
F.G.
designed the research;
S.T.,
E.G.
performed experiments;
O.C.
wrote the wavelet analysis code;
S.T.,
E.G.,
B.Li,
O.C.
analyzed experiments;
S.T.,
B.Li,
F.G.
developed the model;
S.T.,
B.Li,
H.D.-A.,
F.G.
wrote the manuscript;
all authors discussed the results and the manuscript.
}

\competing{We have no competing interests.}

\funding{This work was supported by the European Research Council under the European Union's Seventh Framework Program /ERC consolidator grant agreement 617233 and the LABEX "Who am I?"}

\ack{We warmly thank P. Marcq, S. Yabunaka, N. Graner, J.-F. Rupprecht for critical reading of the manuscript, and T. Das,  F. Gallet, Y. Jiang, U. Schwarz for discussions.}

\disclaimer{NA}

\pagebreak

\bibliographystyle{RS}\bibliography{bibliowaves}

%

\pagebreak

\end{document}